\documentclass[preprint,showpacs,preprintnumbers,nofootinbib,amsmath,amssymb]{revtex4}

\usepackage{amssymb,color}
\usepackage{graphicx}
\usepackage{dcolumn}
\usepackage{bm}

\begin{document}

\title{Dissecting cosmic-ray electron-positron data with Occam's Razor:\\
the role of known Pulsars}

\author{Stefano Profumo$^{1,2}$}
\affiliation{$^1$Department of Physics, University of California, Santa Cruz, CA 95064, USA\\
$^2$ Santa Cruz Institute for Particle Physics, University of California, Santa Cruz, CA 95064, USA
}

\date{\today}

\begin{abstract}
\noindent {\footnotesize We argue that both the positron fraction measured by PAMELA and the peculiar spectral features reported in the total electron-positron flux measured by ATIC have a very natural explanation in electron-positron pairs produced by nearby pulsars. While this possibility was pointed out a long time ago, the greatly improved quality of current data potentially allow to reverse-engineer the problem: given the regions of pulsar parameter space favored by PAMELA and by ATIC, are there known pulsars that explain the data with reasonable assumptions on the injected electron-positron pairs? In the context of simple benchmark models for estimating the electron-positron output, we consider all known pulsars, as listed in the most complete available catalogue. We find that it is unlikely that a single pulsar be responsible for both the PAMELA positron fraction anomaly and for the ATIC excess, although two single sources are in principle enough to explain both experimental results. The PAMELA excess positrons likely come from a set of mature pulsars (age $\sim\times 10^6$ yr), with a distance of 0.8-1 kpc, or from a single, younger and closer source like Geminga. The ATIC data require a larger (and less plausible) energy output, and favor an origin associated to powerful, more distant (1-2 kpc) and younger (age $\sim5\times 10^5$ yr) pulsars. We list several candidate pulsars that can individually or coherently contribute to explain the PAMELA and ATIC data. Although generally suppressed, we find that the contribution of pulsars more distant than 1-2 kpc could contribute for the ATIC excess. Finally, we stress the multi-faceted and decisive role that Fermi-LAT will play in the very near future by (1) providing us with an exquisite measurement of the electron-positron flux, (2) unveiling the existence of as yet undetected gamma-ray pulsars, and (3) searching for anisotropies in the arrival direction of high-energy electrons and positrons.}
\end{abstract}

\pacs{97.60.Gb, 98.70.Sa,95.85.Ry}
\maketitle


\section{Introduction}

Recent measurements of the positron fraction (the fraction of positrons over the sum of positrons and electrons) reported by the PAMELA collaboration \cite{Adriani:2008zr} and of the electron-plus-positron differential energy spectrum reported by ATIC \cite{2008Natur.456..362C} and by PPB-BETS \cite{Torii:2008xu} stirred up great interest both in the cosmic-ray community and among those who work in the field of indirect dark matter searches. 

The Payload for Antimatter Matter Exploration and Light-nuclei Astrophysics (PAMELA) measured a statistically significant trend of increasing positron fraction with energy in the $10\lesssim E_{e^\pm}/{\rm GeV}\lesssim100$ GeV range, at odds with the standard expectation in the context of secondary positron production from interactions of cosmic-ray nuclei with the interstellar gas: if most positrons are of secondary origin, the positron fraction should fall as a function of increasing energy (see e.g. the recent comprehensive stuy of Ref.~\cite{Delahaye:2008ua}). In the past, a similar trend had actually been repeatedly observed, although with much lower statistical significance,  by other experiments, starting  as early as 1969 \cite{1969ApJ...158..771F}. More recent results pointing towards an increasing positron fraction above 10 GeV include the measurement reported in Ref.~\cite{1975ApJ...199..669B} (1975), Ref.~\cite{1987ApJ...312..183M}, Ref.~\cite{1994ApJ...436..769G} (1994), the data reported by HEAT (1997, Ref.~\cite{1997ApJ...482L.191B}) and those quoted by the Alpha Magnetic Spectrometer (AMS) collaboration (2000, Ref.~\cite{Alcaraz:2000bf}). 

The Advanced Thin Ionization Calorimeter (ATIC) recently reported a ``bump'' in the high energy flux of electrons and positrons \cite{2008Natur.456..362C}, in excess of the standard expectation, and marginally in agreement with a statistically less significant excess also visible in the PPB-BETS data reported in Ref.~\cite{Torii:2008xu}\footnote{PPB-BETS measured 84 electron events above 100 GeV \cite{Torii:2008xu}, while ATIC 1724 \cite{2008Natur.456..362C}}. Outside the bump, the ATIC data are in substantial agreement with previous measurements, including those reported by AMS \cite{Alcaraz:2000bf}, BETS \cite{Torii:2001aw}, older PPB-BETS data \cite{Torii:2003ed}, as well as emulsion chambers data \cite{1997AdSpR..19..767N,1980ApJ...238..394N}. The anomalous spectral feature observed in the data from both ATIC flights, and which had never been reported before, is a sharp excess in the electron-positron flux, specifically in the energy range between 300 and 800 GeV \cite{2008Natur.456..362C}. 

In addition to the mentioned experimental results on cosmic-ray electrons and positrons, the atmospheric Cherenkov telescope (ACT) H.E.S.S. recently reported measurements of the flux of very high-energy ($E_{e^\pm}>600$ GeV) electrons, with data extending all the way up to energies around 5 TeV \cite{Collaboration:2008aa}. The data were taken during off-target observations of extragalactic objects. Since no conclusive distinction is possible with ACT's between $e^\pm$ and gamma-rays, the data likely include a contamination from diffuse extra-galactic (and possibly, to some extent, diffuse galactic) gamma rays as well as from residual hadronic cosmic-ray events. The H.E.S.S. data should therefore properly be regarded as {\em upper limits} to the $e^\pm$ flux at large energies, and this is how they will be considered in the present study.

The nature of cosmic-ray positrons have long been believed to be dominantly of secondary origin: the observed cosmic-ray positrons, in that scenario, are the yield of inelastic interactions of (primary and secondary) cosmic-ray protons with the interstellar gas (see e.g. Ref.~\cite{1964ocr..book.....G}, \cite{1977JPhA...10..843G} and \cite{1987MNRAS.228..843D}). Calculations of the cosmic-ray positron and electron flux as well as of the fraction of secondary positrons, assuming several propagation models and diffusion setups, have been carried out extensively over the last 30 years (see e.g. \cite{1982ApJ...254..391P}, \cite{1998ApJ...493..694M}, \cite{Delahaye:2008ua}), with increasing degrees of sophistication, and relying on experimental data of better and better quality . The generic prediction of all these models is a slowly decreasing positron fraction with increasing energy, and of a steeply declining high energy differential primary electron flux (for instance, the conventional and optimized models of Strong et al. \cite{Strong:2004de}, which give a diffuse galactic continuum gamma-ray flux compatible with EGRET data as well as cosmic-ray abundances compatible with a wide set of data, feature injection spectral indexes for high-energy electrons of 2.54 and 2.42, respectively). 

As mentioned above, however, experimental data on cosmic-ray positrons have actually shown for quite a long time a discrepancy with a purely secondary origin for positron cosmic rays. For instance, Ref.~\cite{1989ApJ...342..807B} pointed out almost 20 years ago that, in addition to diffuse sources such as secondary cosmic-ray production and to primary sources, such as electrons re-accelerated in supernova remnant (SNR) shocks, data on the positron fraction and on the $e^\pm$ flux rather conclusively pointed towards the existence of a powerful, nearby source of energetic primary electrons and positrons above 20 GeV. 

As early as 1995, and commenting on data from before 1990, Ref.~\cite{1995PhRvD..52.3265A} stated that ``{\em the measured content of positrons in the total electron flux, [...], at least at high energies [...], is a[n] [...] enigma awaiting an explanation. [...] it is obvious that some other source of positrons is needed}''.  Ref.~\cite{1995PhRvD..52.3265A} also argued for the first time that nearby pulsars might seed a high-energy positron excess, as well as non-standard spectral features in the cosmic-ray electron-positron spectrum. The positron anomaly is therefore nothing {\em qualitatively} new. We argue here, however, that the greatly improved quality of recent data allows us to {\em quantitatively} push previous analyses to the level of associating electron-positron data to existing, observationally well-known galactic objects. This is all the more exciting given the apparently anomalous data reported in the total differential electron-positron flux.

A physical hallmark of energetic ($E_{e^\pm}\gtrsim10$ GeV) electrons and positrons is that they  loose energy very efficiently via two dominant processes: inverse Compton (IC) scattering of background radiation photons, and synchrotron radiative losses. The energy loss rate for both processes goes as the square of the $e^\pm$ energy. Defining the lifetime $t$ of relativistic $e^\pm$ in the interstellar medium as the time it would take for them to cool from an injection energy $E_{e^\pm}$ to rest energy ($E_{e^\pm}=m_e$) if they only lost energy via IC and synchrotron losses, one finds
\begin{equation}
t\sim\ 5\times 10^5\left(\frac{\rm TeV}{E_{e^\pm}}\right)\ {\rm yr}.
\end{equation}
This implies that TeV $e^\pm$ must have been injected not much longer than $\sim10^5$ yr ago. This also implies that, for conventional values of the diffusion coefficient $D$ for the propagation of electrons and positrons, they must have been produced somewhere within $\sqrt{D\times t}\sim$100-500 pc. The efficient energy losses of energetic electrons therefore constrain the sources of high energy cosmic-ray electrons and positrons both in time and in space \cite{1995A&A...294L..41A}. Any interpretation of the reported ``lepton anomalies'' must evidently be compatible with this fundamental physical property of cosmic-ray electron and positron propagation in the interstellar medium (ISM).

Several scenarios have been suggested to explain a deviation of the electron-positron data from the standard cosmic ray secondary positron model (for a recent model-independent analysis see e.g. Ref.~\cite{Serpico:2008te}). A list of these includes:
\begin{enumerate}
\item A cutoff in the primary electron spectrum at large energies (see e.g.~ \cite{1987ICRC....2...88T}). This possibility might help explain the positron fraction data, but is evidently not an explanation to the measured differential $e^\pm$ flux, which would still be dominated, at high energy, by primary cosmic-ray electrons;
\item The production and re-acceleration of secondary electrons and positrons via hadronic processes in giant molecular clouds. For instance, for nominal model parameters, Ref.~\cite{1990ICRC....4..109D} points out that those systems can source a significantly larger positron fraction in excess to the prediction of ordinary diffuse cosmic-ray models, starting at energies around 10 GeV. In this scenario, the results of Ref.~\cite{1999APh....11..429C} indicate that it could be possible to reproduce the PAMELA data below 30 GeV, but no satisfactory explanation to ATIC or to the higher energy bins reported in the PAMELA data seems to be achievable in the context of giant molecular clouds models;
\item Ref.~\cite{1991JPhG...17.1769A} pointed out that $e^\pm$ might be pair-produced near compact gamma-ray sources, for instance as a result of processes involving, in the initial state, the source gamma rays and the optical and UV photons also produced locally by the gamma-ray emitter. Electron-positron pair production could also occur from compact gamma-ray sources in interactions of gamma rays with starlight photons in the interstellar medium, as observed in Ref.~\cite{1991ICRC....2..145M}. This possibility appears, however, to be tightly constrained by current galactic gamma-ray data, and it will be further constrained and quite easily scrutinized by upcoming Fermi data. 
\item Another possible source of primary positrons is the radioactive $\beta^+$ decay of nuclei (such as ${}^{56}$Co) ejected during a supernova explosion \cite{1990ICRC....4...68E, 1979ICRC....1..501L}. Supernova models, however, indicate that the spectrum of such nuclei (and therefore of the positrons produced in the decays) is cut off at energies around 100 GeV, ruling out this scenario as an explanation to the $e^\pm$ spectrum measured by ATIC. Also, the energetics of the process indicate that this mechanism is likely insufficient to produce as large a flux of positrons as to explain the PAMELA positron fraction data too.
\item Pulsars are undisputed sources of electron-positron pairs, produced in the neutron star magnetosphere and, possibly, re-accelerated by the pulsar wind and/or in SNR shocks. Nearby pulsars have since long been considered as very likely dominant contributors to the local flux of high energy electrons and positrons (see e.g. Ref.~\cite{1986ARA&A..24..285T,1988gera.book..480B}). As shown in Ref.~\cite{Hooper:2008kg} and \cite{Yuksel:2008rf}, the recent PAMELA data can rather well be interpreted as originating from one single nearby pulsar, e.g. Geminga, or by the coherent superposition of the $e^\pm$ flux from all galactic pulsars \cite{2001A&A...368.1063Z,Hooper:2008kg} (see also Ref.~\cite{Malyshev:2009tw} for a recent re-assessment of pulsars as the origin of the cosmic-ray lepton anomalies appeared after the present manuscript was first posted);

\item A further possibility that recently received enormous attention is that the excess positrons and electrons originate from the annihilation or decay of particle dark matter (for a possibly incomplete list of related studies appeared before the first version of the present manuscript was released see Ref.~\cite{Hamaguchi:2008ta, Allahverdi:2008jm, Pohl:2008gm, Liu:2008ci,Nath:2008ch,MarchRussell:2008tu,Zhang:2008tb, Pospelov:2008rn,Lattanzi:2008qa,Chun:2008by,Hisano:2008ah,Ibe:2008ye,Taoso:2008qz,Ishiwata:2008qy,Zurek:2008qg,Cholis:2008wq,Morselli:2008uk,Hall:2008qu,Chen:2008qs,Kalinowski:2008iq,Baek:2008nz,Ibarra:2008jk,Pospelov:2008zw,Ponton:2008zv,Hur:2008sy,Hamaguchi:2008rv,Chen:2008md,Fox:2008kb,Bai:2008jt,Ishiwata:2008cv,Yin:2008bs,Feldman:2008xs,Harnik:2008uu,Nomura:2008ru,Cholis:2008qq,Bringmann:2008jn,Masip:2008mk,Cirelli:2008pk,Cirelli:2008jk,Nelson:2008hj,Pospelov:2008jd,Huh:2008vj,Cholis:2008hb,Barger:2008su,Bergstrom:2008gr,Bae:2008sp,Chen:2008xx}; further studies appeared between the first and the present version of this manuscript include \cite{Grajek:2008pg, Huh:2008ez, Bi:2009md, Cui:2009xq, Gogoladze:2009kv, Cao:2009yy, Guo:2009aj, Takahashi:2009mb, Shepherd:2009sa, Nezri:2009jd, Chen:2009mf, Meade:2009rb, Mardon:2009rc, deBoer:2009rg, Brandenberger:2009ia, Banks:2009rb, Kribs:2009zy, Chang:2009dh, Kyae:2009jt, Hamaguchi:2009sz, Phalen:2009xw, Chen:2009dm, Chen:2009iu, Hooper:2009fj, Goh:2009wg, Ibe:2009dx, Cheung:2009qd, Allahverdi:2009ae, Bae:2009bz, Bringmann:2009ip, Cheung:2009si, Shirai:2009kh, Bi:2009uj, Ishiwata:2009vx, Frampton:2009yc, Finkbeiner:2009mi, Cassel:2009pu, Roszkowski:2009sm, Chen:2009ew, Rothstein:2009pm, Ishiwata:2009pt, Zant:2009sv, Brun:2009aj, McDonald:2009cc, Lingenfelter:2009kx, Barger:2009yt, Gogoladze:2009gi, Morrissey:2009ur, Arvanitaki:2009yb, Davoudiasl:2009dg, Kuhlen:2009is, Kawasaki:2009nr}). As opposed to all of the other possibilities mentioned above, a dark matter interpretation invokes an entity whose fundamental particle physics nature has yet to be unveiled, and whose existence in the form of a particle with a mass in the GeV-TeV range is yet to be demonstrated. In addition to this, there are at least three general weaknesses to the dark matter interpretation of the positron fraction and of the features detected in the high energy electron-positron flux:
\begin{enumerate}
\item the annihilation or decay products of dark matter generically include both leptons and hadrons. Unless the dark matter model is tuned in such a way as to make it ``lepto-philic'' \cite{Fox:2008kb}, the hadronic products will typically overproduce both gamma rays and antiprotons, in contrast with experimental data (e.g. \cite{Bergstrom:2006tk});
\item the annihilation rate required to explain the cosmic-ray data under discussion is typically a couple orders of magnitude larger than what expected if the dark matter is cosmologically produced through thermal freeze-out in the right amount. This implies a further model-building intricacy. In addition, low-velocity enhancements, which would lead to a burst of annihilations in the first collapsed halos \cite{Profumo:2006bv}, are generically very tightly constrained by observations of the cosmic diffuse background radiation \cite{Kamionkowski:2008gj};
\item The large amount of energetic positrons and electrons injected in our own Galaxy and in any dark matter structure in the Universe would yield secondary emission at all wave-lengths (for a recent comprehensive analysis see Ref.~\cite{Bergstrom:2008ag}): this includes radio signals from synchrotron losses, X-rays and gamma-rays from inverse Compton scattering off the cosmic microwave background and other background radiation photons, as well as from bremsstrahlung, neutral pion decay and internal bremsstrahlung radiation \cite{Beacom:2004pe}. Tight constraints exist at each one of these wavelengths, at all dark matter halo scales and distances, ranging from local dwarf galaxies to distant clusters (a very incomplete list of references on multi-wavelength constraints on dark matter annihilation includes e.g. \cite{Totani:2004gy,Colafrancesco:2005ji, Profumo:2005xd,Profumo:2006hs,Colafrancesco:2006he,Profumo:2008fy,Jeltema:2008ax,Jeltema:2008hf,Jeltema:2008vu,PerezTorres:2008ug, Essig:2009jx, Natarajan:2007ja, Colafrancesco:2004sp, Colafrancesco:2009jq, Borriello:2008dt, Borriello:2008gy, Caceres:2008dr, Regis:2008ij, Hooper:2008zg, Taoso:2007qk, Hooper:2007gi}).
\end{enumerate}
\end{enumerate}

Recent studies, appeared after this manuscript was first released, discuss further possible explanations to the cosmic ray ``lepton anomalies''. These include: a gamma-ray burst \cite{Ioka:2008cv}, inelastic collisions between high energy cosmic rays and background radiation photons in the astrophysical sources where cosmic rays are accelerated \cite{Hu:2009bc, Diehl:2009kp}, inhomogeneities in the spatial distribution of sources of cosmic-ray leptons \cite{Shaviv:2009bu}, hadronic processes producing secondary positrons in the very same sites where primary cosmic rays are accelerated \cite{Blasi:2009hv, Blasi:2009bd} and s scenario where one (or more) recent suernova event(s) occurred in gas clouds close to the Solar System \cite{Fujita:2009wk}.

Of all possibilities listed above, we focus in the present study on the one which we believe -- and intend to argue with the present analysis -- is the best motivated: nearby, known galactic pulsars. On the one hand, pulsars are predicted to produce $e^\pm$ pairs with a spectrum that is suitable in principle to explain both the data on the positron fraction and, possibly, the overall electron-positron differential flux. On the other hand, pulsars are well known and well studied objects, and catalogues of galactic pulsars exist \cite{2005AJ....129.1993M,2005yCat.7245....0M}, off of which we can compare theoretical speculations on the most plausible sources for the observed $e^\pm$ fluxes with actual observational data.

Unlike in the dark matter hypothesis, the pulsar scenario does not require the introduction of an extra, exotic component to explain the data. This goes well with the celebrated quote of the 14th-century English logician and Franciscan friar, William of Ockham: ``{\em Entia non sunt multiplicanda praeter necessitatem}'': there is no need to introduce extra entities beyond necessity.

The scope of the present analysis is to  ``dissect'' the PAMELA and ATIC data in the spirit of (or with) Occam's razor: we will pinpoint which regions of the pulsar parameter space (to be defined in the next section) are favored by experimental data; we will then turn to observations, and see if those parameter space regions correspond to existing astrophysical objects, and if those objects are predicted to have an $e^\pm$ energy output in line with what is needed to explain the cosmic ray data. We shall argue that the recent positron and electron data are certainly very intriguing, but that they are {\em not anomalous}, since a very natural interpretation exists in terms of known, existing astrophysical objects. We will also outline the major uncertainties that enter the process of reverse-engineering cosmic-ray lepton data to infer their possible origin. Among these, uncertainties in the pulsars' age and location at the point in time when the bulk of the high-energy lepton pairs where injected play an extremely relevant role.

In summary, highlights and novelties presented in this study include:
\begin{itemize}
\item accounting for all known, existing pulsars, as listed in the ATNF catalogue \cite{2005AJ....129.1993M}: this approach allows us to (i) understand, qualitatively, which regions of the pulsar age-versus-distance parameter space contribute in given ranges of the cosmic ray electron-positron spectrum (see sec.\ref{sec:all} and fig.~\ref{fig:distant} and \ref{fig:distageall}), (ii) analyze, through the determination of the relevant age-distance range, where the pulsar catalogue incompleteness can affect predictions for the high-energy cosmic-ray electron-positron spectrum, and thus assess where and how newly discovered gamma-ray pulsars can impact our understanding of the origin of high-energy cosmic-ray electrons (see e.g. sec.\ref{sec:glast} and fig. \ref{fig:CTA1}), (iii) address how the hereby proposed ``reverse-engineering'' approach (and in general any attempt at pinpointing a local high-energy electron-positron source as the main contributor to the leptonic cosmic-ray fluxes measured at Earth) is affected by uncertainties in the determination of the pulsars' age and distance, due to e.g. to trapping of the $e^\pm$ in the pulsar wind nebula, and to pulsar kick velocities (see fig.~\ref{fig:bestfit_distage}), (iv) estimate, by considering all ATNF pulsars, the theoretical uncertainties built in several current ``benchmark'' models for the $e^\pm$ energy injection from pulsars, such as those described in the next section (see fig.~\ref{fig:output} and \ref{fig:modelsagedist});
\item systematically analyzing and quantifying the role of uncertainties related to cosmic-ray energy losses and diffusion modeling (see e.g. fig.\ref{fig:single_hard}-\ref{fig:pamela}), and quantifying how these uncertainties impact the identification of pulsars that can source the positron fraction data and/or the $e^\pm$ spectral features; 
\item pointing out that there are regions of the pulsar age-distance parameter space that are more likely to contribute to the detected lepton anomalies (see e.g. fig.\ref{fig:atic} and tab.~\ref{tab:combination}). This point is relevant to the possibility of carrying out anisotropy studies, since it provides e.g. target directions for the possible detection of a dipolar asymmetry in the CRE arrival directions. In addition, as we point out with fig.\ref{fig:glast}, Fermi data may soon tell us whether one pulsar or a combination of pulsars are at the origin of the high-energy end of the cosmic-ray electron-positron spectrum, and has the potential to actually detect the mentioned anisotropies;
\item theoretically studying the role of leaky-box boundary conditions  to the widely-used analytical expression \cite{1995PhRvD..52.3265A} employed to calculate the spectrum, at Earth, of cosmic-ray electrons and positrons injected by a pulsar in the burst-like approximation; this physically means that we provide an estimate of the fraction of electrons and positrons injected by a given pulsar at a given point in time that freely escapes the galactic diffusive region, as a function of the pulsar and of the diffusion model parameters (see the last few paragraphs of sec.~\ref{sec:models} and fig.~\ref{fig:spherical}).
\end{itemize}

It goes without saying that the quest for the fundamental nature of particle dark matter, in this case indirect detection, needs to undergo the perhaps painful, or prosaic, process of deep scrutiny of any ``mundane'', standard interpretation of experimental data, including known astrophysical processes. The present study is a modest attempt in this direction. An important point we wish make at the end of this study is that the Fermi gamma-ray space telescope will play a decisive role, in the very near future, to elucidate the nature of the $e^\pm$ sources, as well as to inform us on what particle dark matter can or cannot be \cite{Baltz:2008wd}.

The outline of this study is as follows: in the next section we describe the processes that can seed $e^\pm$ pair production in pulsars, estimate the $e^\pm$ energy output and injection spectrum, and discuss the subsequent particle propagation, including a novel analytical formula to account for leaky-box boundary conditions in the propagation setup. Sec.~\ref{sec:nearby} gives a few examples of nearby pulsars that could contribute to the $e^\pm$ flux observed locally at the level of the measurements reported by ATIC and PAMELA. The following section addresses the question of whether those measurements can be explained by a {\em single} nearby source. Sec.~\ref{sec:pamela} and \ref{sec:atic} are then devoted to outlining the pulsar parameter space favored by the PAMELA and ATIC data, respectively. Sec.~\ref{sec:onemulti} contrasts a scenario where the measured $e^\pm$ originate from a single source to one where they stem from multiple pulsars in well defined parameter space regions. We carry out a self-consistency check of our models in sec.~\ref{sec:all}, where we compute the $e^\pm$ summing over the full galactic pulsar catalogue. Finally, we outline in sec.~\ref{sec:glast} the decisive role of the Fermi telescope in understanding the $e^\pm$ data: (1) to provide an exquisitely detailed measurement of the $e^\pm$ spectrum from 20 GeV to 1 TeV, and (2) to discover potentially relevant new gamma-ray pulsars, and , and (3) to search for anisotropies in the arrival direction of high-energy electrons and positrons. Sec.~\ref{sec:concl} summarizes and concludes.

\section{Benchmark Models for the Production of Electron-Positron Pairs from Pulsars}\label{sec:models}

Pulsars are known to be potentially very powerful sources of primary electrons and positrons, pair-produced in the neutron star magnetosphere.
Two classes of mechanisms of pulsar relativistic particle acceleration (and associated gamma-ray and $e^\pm$ pair-production) have been widely discussed in the literature, corresponding to the regions of the magnetosphere where the particle acceleration takes place: the {\em polar cap} model (see e.g. Ref.~\cite{1975ApJ...196...51R,1983ApJ...266..215A,1982ApJ...252..337D,1996A&AS..120C.107D,1985Ap&SS.109..365R,1995AuJPh..48..571U}) and the {\em outer gap} model (\cite{1986ApJ...300..500C,1996ApJ...470..469R,2002BASI...30..193H}). We will not review those mechanisms in any detail here. In short, in both models very high energy electrons are accelerated by electric fields in the pulsar magnetosphere. The electrons, in turn, synchrotron-radiate gamma rays with energies large enough to pair-produce electron-positron pairs in the intense magnetic fields of the magnetosphere, and/or through scattering off of intervening photon fields, such as the thermal X-rays produced by the pulsar itself. We only mention here that, in the context of the outer gap model, the threshold for $e^\pm$ pair-production from the scattering off of the pulsar thermal X-rays onsets when the ratio of the size of the outer gap versus the radius of the light cylinder is less than one (for details see Ref.~\cite{1994AIPC..304..116C} and \cite{1997ApJ...487..370Z}). Pulsars that fulfill this condition are indicated as ``{\em gamma-ray pulsars}''. In terms of the pulsar magnetic field at the neutron star surface  $B_{12}$, in units of $10^{12}$ G, and of the pulsar period $P$ in seconds, Ref.~\cite{1997ApJ...487..370Z} finds that gamma-ray pulsars correspond to those objects that fulfill the condition
\begin{equation}
g=5.5\ P^{26/21}\ B_{12}^{-4/7}\ <\ 1.
\end{equation}

A crucial parameter for the present study is the estimate of the total energy that a pulsar radiates in electron-positron pairs that actually diffuse in the interstellar medium (ISM).
To estimate the total energy output in $e^\pm$ from pulsars with rotation period $P/{\rm s}$ (corresponding to a rotational frequency $\Omega=2\pi/P$ Hz) and a period derivative $\dot P$ , we adopt here the classic magnetic dipole pulsar model outlined e.g. in Ref.~\cite{1969ApJ...157.1395O}, and reviewed e.g. in \cite{1986bhwd.book.....S}. In this scheme, pulsars consist of a rotating neutron star featuring a dipolar magnetic field which is not aligned with the rotation axis. The neutron star, with a moment of inertia $I\sim1.0\times 10^{45}\ {\rm g}\ {\rm cm}^2$ and angular frequency $\Omega$ is assumed to lose rotational energy via magnetic dipole radiation, and to follow a braking law of $\dot\Omega\propto-\Omega^3$. This can be integrated to yield (in the limit of a mature pulsar, i.e. one with a rotation period much longer than its initial one) a characteristic age $T=\Omega/(2\dot\Omega)$. 

Notice that this estimate of the pulsar age does not necessarily correspond to the {\em actual} pulsar age (some of the pulsar ages in the ATNF catalogue obtained in this way, in fact, turn out to be even larger than the age of the Universe!). This problem is particularly severe for young pulsars with ages between $10^4$ and $10^5$ yr. Ref.~\cite{1999ApJ...523L..69T} and \cite{2001ApJ...560..371K}, for instance, pointed out that the pulsar J1811-1925 associated to the SNR G11.2-0.3, whose age is inferred from the likely association with the supernova reported by Chinese astronomers in A.D. 386, has an actual age which is more than one order of magnitude smaller than the characteristic age derived from timing measurements, which is of the order of 24,000 yr. From a semi-analytical standpoint, in the model of Ref.~\cite{FaucherGiguere:2005ny}  the median over-estimate of the pulsar age obtained from timing measurements is of 60,000 yr, which gives a yard-stick to estimate whether for a given set of pulsars this uncertainty is or not relevant. 

In addition to the uncertainty in the actual pulsar age, another effect should be taken into account when considering a burst-like injection of $e^\pm$ from pulsars: before diffusing in the ISM, the electrons and positrons produced by the pulsar are trapped in the pulsar wind nebula or in the supernova remnant that envelopes it. This time-lag is of the order of $10^4...10^5$ yr (for a review on pulsar wind nebulae we refer the reader to Ref.~\cite{Gaensler:2006ua}). This time-delay also implies that pulsars relevant for the present discussion are {\em mature} pulsars, with typical ages in excess of $10^5$ yr -- younger pulsars likely having not expelled yet the produced electron-positron pairs in the ISM. 

We outline below here four possible schemes to estimate the total energy injected in the ISM by pulsars in the form of electron-positron pairs, and evaluate the corresponding ``theoretical uncertainty'' by evaluating the output of each model for all pulsars listed in the most complete available catalogue. 

The spin-down power of a pulsar corresponds to the pulsar's rotational energy dissipated in dipolar electro-magentic radiation, and is given by $\dot E\equiv I\Omega\dot \Omega$. The dipole magnetic field strength at the stellar surface, again in the classic pulsar dipole model of Ref.~\cite{1969ApJ...157.1395O}, reads
\begin{equation}
B=3.2\times 10^{19}\sqrt{P\times \dot P}\ {\rm G}.
\end{equation}
We indicate with $f_{e^\pm}$ the fraction of the rotational energy (integrated spin-down power) which is eventually dissipated in $e^\pm$ pairs diffusing in the ISM. We can naively estimate the energy output as follows: 
The general solution to the pulsar rotational velocity, in the regime where the pulsar looses energy dominantly via dipole radiation, can be cast as
\begin{equation}
\Omega(t)=\frac{\Omega_0}{\left(1+t/\tau_0\right)^{1/2}},
\end{equation}
where $\Omega_0$ is the initial spin frequency of the pulsar, and, indicating with $R_s$ the neutron star radius, 
\begin{equation}
\tau_0=\frac{3c^3I}{B^2R_s^6\Omega_0^2}
\end{equation}
is a characteristic ``decay time'' which, incidentally, cannot be directly derived from pulsar timing measurements, and is usually assumed to be, for nominal pulsar parameters, $\sim10^4$ yr \cite{1995A&A...294L..41A}. For known mature pulsars, which could contribute to the observed $e^\pm$ flux because the produced particles are no longer trapped and lose energy in the nebula \cite{1996ApJ...459L..83C}, we approximate the total energy output as 
\begin{equation}
  E_{\rm out}\ =\ f_{e^\pm}\times I\times \int{\rm d}t \ \Omega \dot\Omega \approx \frac{f_{e^\pm}}{2}I\Omega_0^2
\end{equation}
Now, since 
\begin{equation}
\Omega_0^2\simeq\Omega^2\frac{t}{\tau_0},
\end{equation}
we have, for $t=T$, i.e. the age of the pulsar,
\begin{equation}
E_{\rm out}\ [{\rm ST}]=\frac{f_{e^\pm}}{2}I\Omega^2\frac{T}{\tau_0}=f_{e^\pm}(I\Omega\dot\Omega)\left(\frac{\Omega}{2\dot\Omega}\right)\frac{T}{\tau_0}=f_{e^\pm}\dot E\frac{T^2}{\tau_0}.
\end{equation}
In this naive scenario, the pulsar output in $e^\pm$ is then completely fixed by the rotation period and its derivative through the derived values of the spin down power $\dot E$ and typical age $T$, once the efficiency factor for $e^\pm$ production is taken into account. We indicate hereafter this model as the ``ST'' model. Unless otherwise specified we assume $f_{e^\pm}=0.03$, i.e. 3\% of the rotational energy gets converted into ejected $e^\pm$ pairs. This is a rather conservative assumption, along the lines of what assumed e.g. in Ref.~\cite{Hooper:2008kg}, i.e. $f_{e^\pm}\sim$ few \%. A quantitative discussion of plausible values for $f_{e^\pm}$ was recently given in Ref.~\cite{Malyshev:2009tw}. We shall not review their discussion here, but Ref.~\cite{Malyshev:2009tw} argues (see in particular their very informative App.~B and C) that in the context of a standard model for the pulsar wind nebulae, a reasonable range for $f_{e^\pm}$ falls between 1\% and 30\%.

An alternative scheme is the model for the $e^\pm$ pulsar energy output formulated by Harding and Ramaty \cite{1987ICRC....2...92H} (hereafter, the HR model), where the main assumption is that the pulsar luminosity in $e^\pm$ is proportional, via a factor $f_{HR}$, to the gamma-ray output (the high-energy gamma rays and $e^\pm$ are assumed to follow the same spectral index):
\begin{eqnarray}
& L_{e^\pm}=f_{HR}\times L_\gamma,\quad {\rm where} &\\
\nonumber & L_\gamma\simeq7.1\times 10^{33}(E/{\rm GeV})^{-\alpha}\ B_{12}\ P^{-1.7}\ {\rm ph}\ {\rm s}^{-1}\ {\rm GeV}^{-1}&
\end{eqnarray}
The expression given above is, incidentally, in agreement with data from the Crab and from the Vela pulsar, as pointed out in Ref.~\cite{1987ICRC....2...92H}. HR then assume that positrons are produced with gamma rays in the cascade processes described above. Particle Monte Carlo simulations indicate that in electro-magnetic showers the ratio of $e^\pm$ pairs to gamma-rays $f_{HR}$ lies between 0.2 and 0.5 in the energy range between 1 and 10 GeV \cite{1982ApJ...252..337D}. Substituting the pulsar age (in s) and magnetic field for the rotation period (also in s) according to
\begin{equation}\label{eq:substitution}
P=\frac{B_{12}\sqrt{2T}}{3.2\times 10^7},
\end{equation} 
and integrating over time, the total $e^\pm$ energy output in the HR can be estimated as
\begin{equation}
E_{\rm out}\ [{\rm HR}]=1.3\times 10^{46}\ f_{HR}\ B_{12}^{-0.7}\ \left(\frac{T}{10^4\ {\rm yr}}\right)^{0.15}\ {\rm erg}.
\end{equation}

In the outer gap model of Chi, Cheng \& Young (CCY) \cite{1996ApJ...459L..83C}, the $e^\pm$ output can be assessed by taking the product of the total pair production rate in their setup (to which we refer the reader for additional details)
\begin{equation}
\dot N_{e^\pm}\simeq3.4\times 10^{37}\ g\ B_{12}^{10/7}\ P^{-8/21}\ {\rm s}^{-1}
\end{equation}
and of the typical electron injection energy
\begin{equation}
\tilde E\simeq1.1\times 10^6\ B_{12}^{4/7}\ P^{-76/21}\ {\rm eV}
\end{equation}
which, after substitution of Eq.~(\ref{eq:substitution}) and after integration over time, gives:\begin{equation}
E_{\rm out}\ [{\rm CCY}]=7.6\times 10^{48}\ \zeta\ B_{12}^{-9/7}\ \left(\frac{T}{10^4\ {\rm yr}}\right)^{-5/14}\  {\rm erg},
\end{equation}
where $\zeta\sim 0.3$ parameterizes the escape efficiency from the light cylinder.

A similar but slightly different construction was proposed by Zhang and Cheng (ZC) \cite{2001A&A...368.1063Z}: here, following their analysis and the procedure outlined above, we arrive at an energy output, for the ZC model, of
\begin{equation}
E_{\rm out}\ [{\rm ZC}]=4.8\times 10^{48}\ B_{12}^{-2}\ \left(\frac{T}{10^4\ {\rm yr}}\right)^{-1}\  {\rm erg}.
\end{equation}

We now proceed to assessing the $e^\pm$ output for pulsars appearing in the Australia Telescope National Facility (ATNF) pulsar catalogue\footnote{{\tt http://www.atnf.csiro.au/research/pulsar/psrcat/}} \cite{2005AJ....129.1993M}. The ATNF sample provides the most exhaustive and updated list of known pulsars, including on only radio pulsars, but also those objects that were detected only at high energy (e.g. X-ray pulsars and soft gamma-ray repeaters).  The total number of objects we use here (essentially all ATNF pulsars for which measurements of the pulsar period, its derivative and the distance are available) is 1789, of which 266 objects feature $g<1$ (``gamma-ray pulsars'' according to the definition above). The first point we wish to assess here is how large a range of predictions the four benchmark models outlined above give when applied to the entire ATNF pulsar catalogue. This will provide us with a rule of thumb on how uncertain predictions for the {\em normalization} of the contribution of known pulsars to the flux of local cosmic-ray electrons and positrons are.

\begin{figure*}
\includegraphics[width=17.cm,clip]{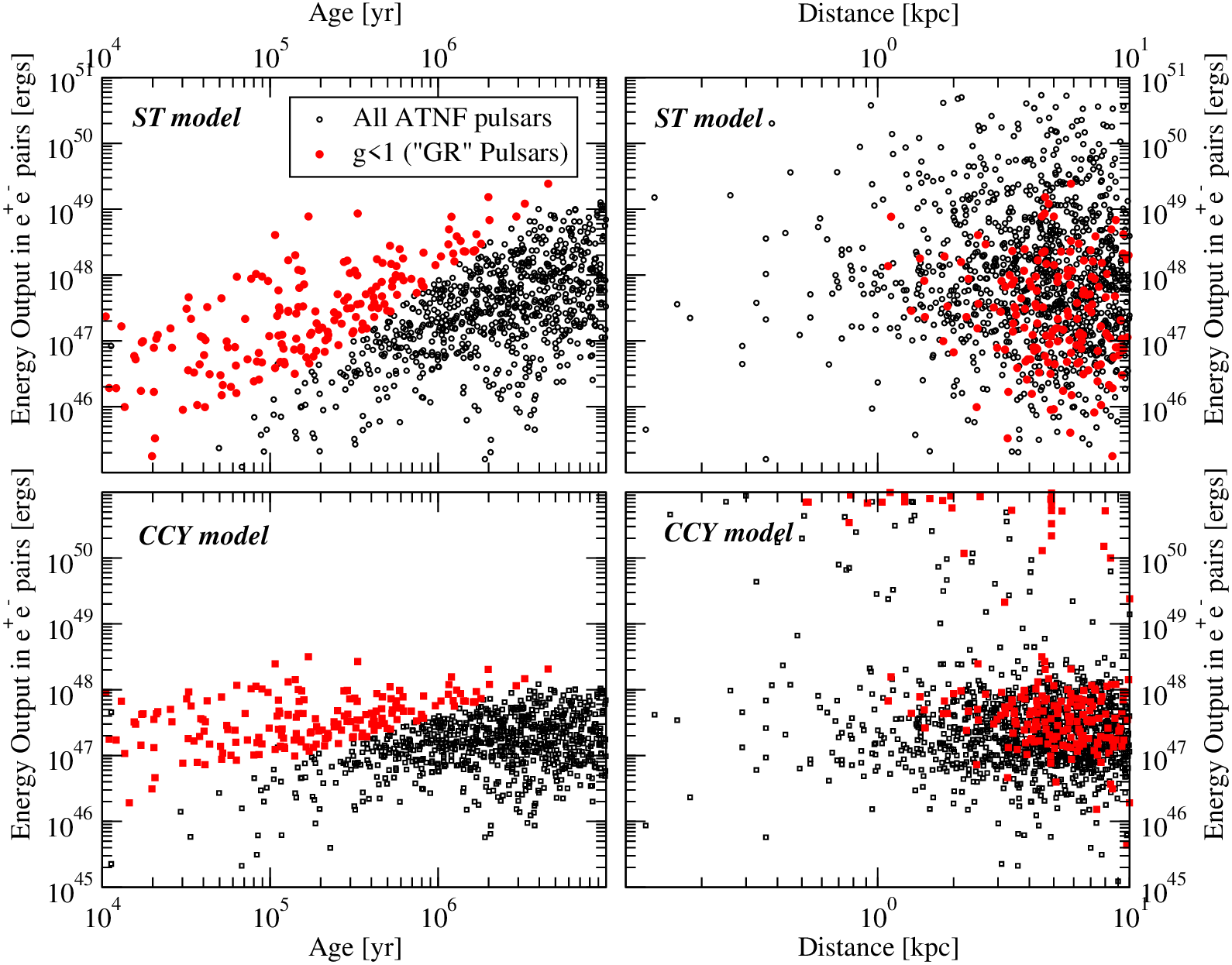}
\caption{\label{fig:output} Scatter plot in the energy output in $e^\pm$, in ergs, versus age (left panels) and versus distance (right panels), for the ST model and for the CCY model. The black circles indicate all pulsars in the ATNF catalogue, while the red dots indicate ``gamma-ray'' pulsars ($g<1$).}
\end{figure*}

In  fig.~\ref{fig:output} we show the estimated pulsar energy output in $e^\pm$ for the entire ATNF sample, for the ST model (upper panels) and for the CCY model (lower panels). The two panels to the left indicate the pulsar position in the age versus $e^\pm$ output plane (the energy output in ergs), while on the right we show the distance versus output plane. Red points highlight those pulsars whence a contribution from the outer gap is expected (``gamma-ray'' pulsars, $g<1$).

The first feature we observe in our results is that for pulsars with ages $10^5<T/{\rm yr}<10^6$ the model predictions fall between $10^{47}$ and $10^{48}$ erg. We remark here that these values are in agreement with older estimates, see e.g. \cite{1995PhRvD..52.3265A} and wit the estimates quoted in more recent studies \cite{Yuksel:2008rf,Hooper:2008kg, Malyshev:2009tw}. We notice that the relevant energy output range, quite remarkably, does not critically depend on the choice of the model for the estimate of the $e^\pm$ output. The spatial distribution of pulsars places the $g<1$ outer gap candidates at around a kpc or so, with a larger scatter in the ST model (with young pulsars contributing lower outputs and older pulsars larger outputs).

\begin{figure*}
\mbox{\includegraphics[width=8.cm,clip]{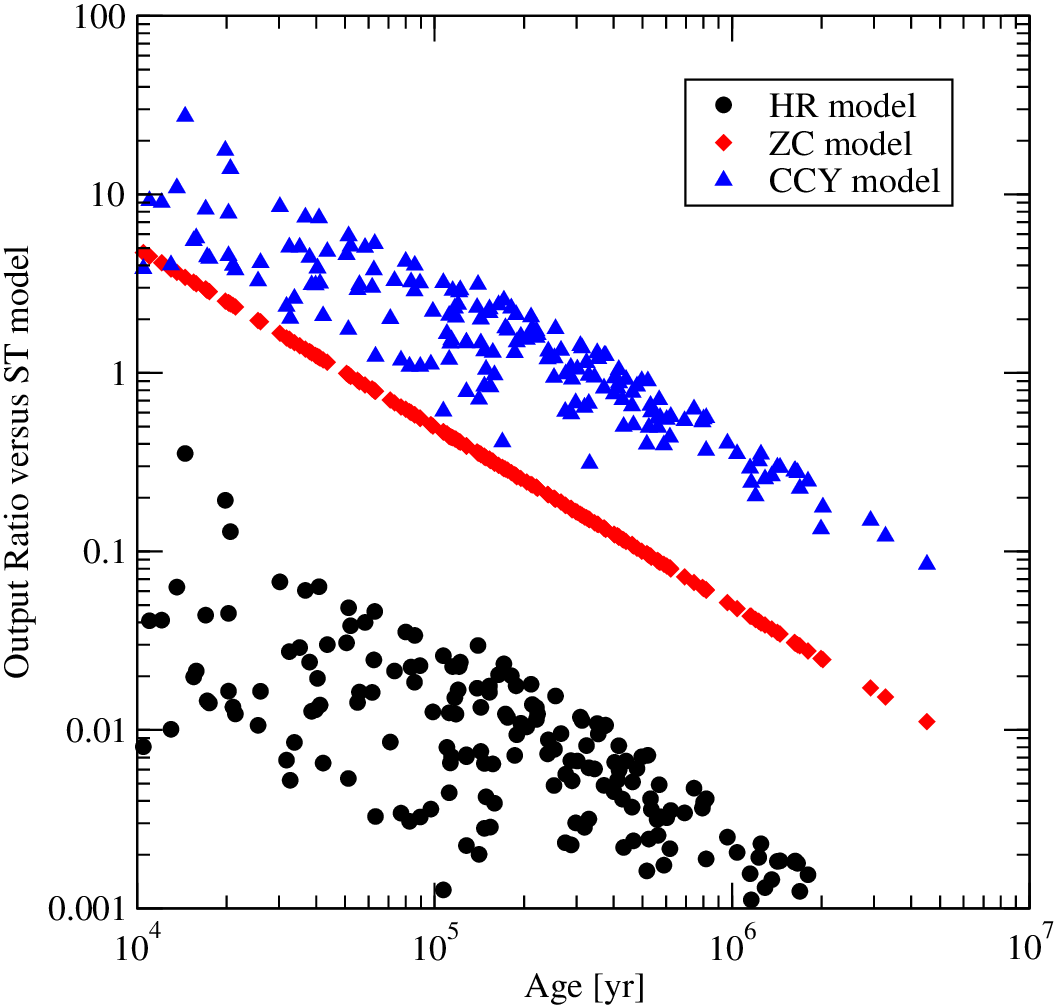}\qquad\includegraphics[width=8.cm,clip]{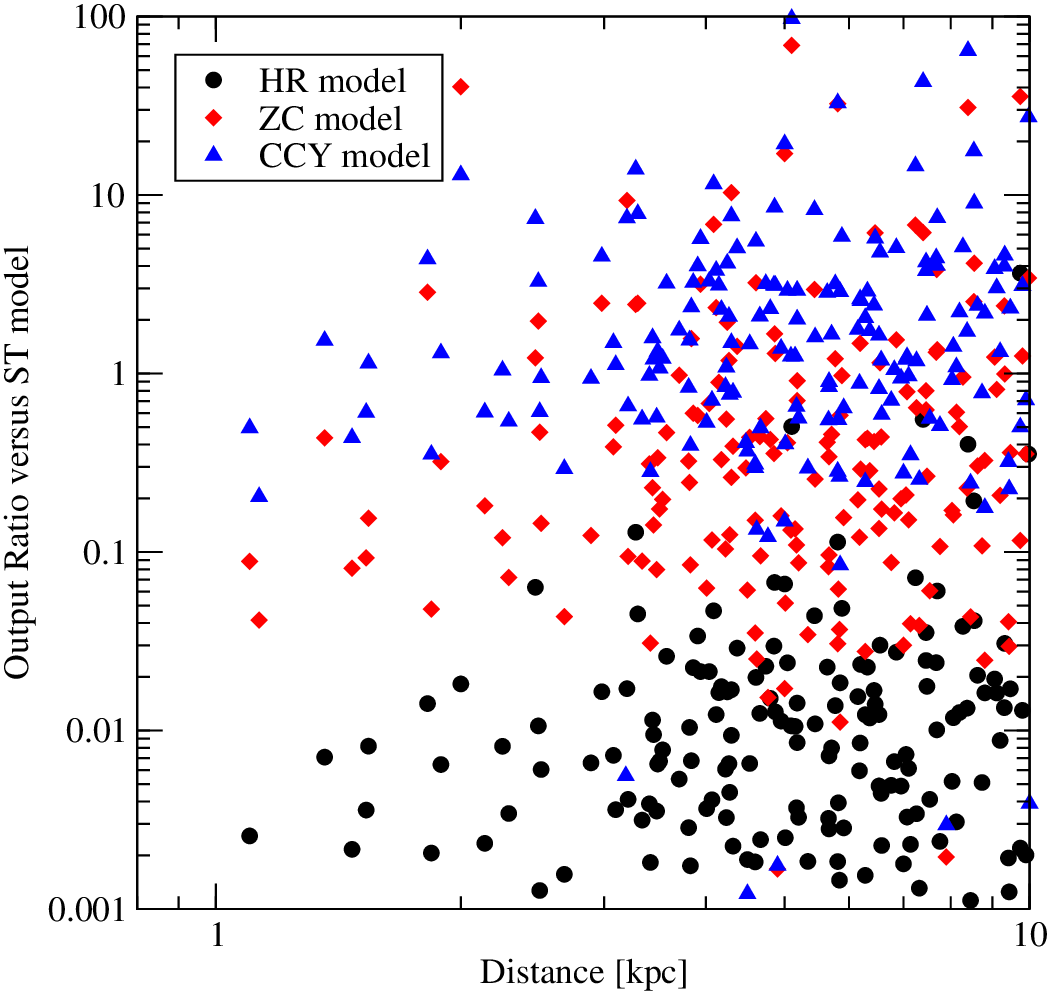}}
\caption{\label{fig:modelsagedist}The ratio of the predicted $e^\pm$ energy output from ``gamma-ray'' ($g<1$) pulsars for the HR, ZC and CCY models over the naive prediction (ST model), as a function of age (left panel) and of distance (right panel).}
\end{figure*}

In the following fig.~\ref{fig:modelsagedist} we illustrate the ratio of the three HR, CCY and ZC model predictions for the pulsar $e^\pm$ output over the ST prediction, as a function of age (left) and of distance (right). First, we notice that the HR model tends to predict a significantly lower output compared to other scenarios, as already noticed in the past \cite{2001A&A...368.1063Z}. Secondly, compared to the CCY and ZC models, the ST model predicts a much larger output for mature pulsars versus younger pulsars. In particular, it is easy to ascertain that the ratio of the ZC to the ST model scales exactly as $T^{-1}$, which quantifies the bias for the ST prediction towards older pulsars. No significant correlations with distance are observed for the ratios among the various models.

Once the total output is specified, the injected electron-positron spectrum is fully defined by its spectral shape. It is well known that $e^\pm$ escaping the pulsar wind cannot be accelerated to arbitrarily high energies, and a cut-off is expected, depending on the pulsar environment, in the energy range around the TeV scale. The precise position of the cutoff is quite uncertain, and it also depends critically on the pulsar age and on the magnetic fields in the pulsar wind nebula (see e.g.~\cite{Busching:2008zzb, Buesching:2008hr}). Although physically there will be an upper limit to the energy of the $s^\pm$ escaping the pulsar, this would (i) introduce one more parameter in the analysis we present below, and make it easier, in general, to accommodate the observed lepton anomalies by tuning it, and (ii) for the energies of interest and pulsars under consideration, the precise location of the cutoff is often not relevant (for instance because the pulsar is old enough that the maximal energy of $e^\pm$ propagating from it to the Earth is well below the injection spectrum cutoff). We therefore only illustrate the impact of a cutoff in the injection spectrum in a few selected cases below -- if not specified otherwise, we will not assume any injection spectrum cutoff in what follows.

Below the spectral cutoff in the injected flux of $e^\pm$, it is reasonable to assume a power law,
\begin{equation}
\frac{{\rm d}N_{e^\pm}}{{\rm d}E_{e^\pm}}\propto E_{e^\pm}^{-\alpha}.
\end{equation}
The normalization can then be extracted by a simple integration above the relevant energy threshold, conventionally taken to be $\sim1$ GeV. Information on the spectral index $\alpha$ can be inferred from actual multi-wavelenght pulsar observations. Secondary synchrotron radiation at radio frequencies indicate, for the $e^\pm$ population of e.g. pulsar wind nebulae, spectral indexes ranging from 1 to 1.6 \cite{1988ApJ...327..853R}. Larger values of $\alpha$, or a steepening of the spectrum at larger energies, are implied from the detected X-ray emission from the same objects. In addition, e.g. in the HR model the spectral index of gamma-ray from pulsars should be close to the spectral index of the injected $e^\pm$. EGRET observations of galactic pulsars point to fairly hard gamma-ray spectra in the 0.1 to 10 GeV range, loosely between 1.4 and 2.2 \cite{Thompson:1994sg,Fierro:1995ny}. For all these reasons, we shall consider here the rather liberal range of values $1.4<\alpha<2.4$.

In addition to the electron-positron spectrum, a potentially important impact on the measured $e^\pm$ flux on Earth from pulsar physics stems from the finite time during which the energy output estimated above is delivered in the inter-stellar medium. As shown e.g. in fig.~4 of Ref.~\cite{1995PhRvD..52.3265A}, different decay time scales, or different models for the $e^\pm$ injection time profile, impact the spectrum of the $e^\pm$ produced in pulsars. Specifically, a tail at high energy extends in the $e^\pm$ spectrum beyond the cutoff energy set by inverse Compton and synchrotron losses. There is no reason to believe that there exist one unique injection spectrum for galactic pulsar emission of electrons and positrons. Including this effect is ultimately important (and technically very easy, once the form of the light curve is assumed), but it would only add an extra parameter (per pulsar) to the present discussion. As for the injection spectrum cutoff, adding more parameters would only make it easier to demonstrate the point we want to make: existing known pulsars provide a natural explanation to the positron fraction and to the electron-positron differential spectrum data. Therefore, we shall hereafter assume a bursting $e^\pm$ injection, i.e. that the duration of the emission is much shorter than the travel time from the source \cite{Hooper:2008kg}, which, incidentally, is an appropriate approximation for almost all pulsars we consider here.

After production, the $e^\pm$ emitted from a galactic pulsar diffuse and loose energy prior to arrival at the Earth's atmosphere. These processes have been described in the standard diffusion approximation (i.e. neglecting convection) by a familiar diffusion equation \cite{1964ocr..book.....G}, which, for local sources (in a leaky box type scenario, where the height of the diffusion region is larger than the sphere where the contributing sources lies) can be assumed to have spherical symmetry, and reduces to the form \cite{1995PhRvD..52.3265A}:
\begin{equation}\label{eq:diffusion}
\frac{\partial f}{\partial t}=\frac{D(\gamma)}{r^2}\frac{\partial}{\partial r}r^2\frac{\partial f}{\partial r}+\frac{\partial(P(\gamma)\ f)}{\partial\gamma}+Q.
\end{equation}
In the equation above, $f(r,t,\gamma)$, with $\gamma=E_{e^\pm}/m_e$, is the energy distribution function of particles at instant $t$ and distance $r$ from the source; 
\begin{equation}
P(\gamma)=p_0+p_1\gamma+p_2\gamma^2
\end{equation}
indicates the energy loss term (which for our purposes can be very accurately approximated by the last term only, describing the dominant inverse Compton and Synchrotron losses, $P(\gamma)\simeq p_2\gamma^2$ at high energy); $D(\gamma)\propto\gamma^\delta$ is the energy-dependent diffusion coefficient, which is here assumed to be independent of $r$ (i.e. we assume a homogeneously diffusive medium; for a discussion of the role of a non-homogeneous medium, see App.~C in Ref.~\cite{Malyshev:2009tw}).

As mentioned above, the dominant $e^pm$ energy losses at energies at or above $E_{e^\pm}\gtrsim10$ GeV are (i) inverse Compton scattering off photon backgrounds such as the cosmic microwave background (CMB) and photons at optical and IR frequencies, and (ii) synchrotron losses in the ISM magnetic fields. Following Ref.~\cite{1995PhRvD..52.3265A}, we estimate:
\begin{equation}
p_2=5.2\times 10^{-20}\frac{w_0}{1\ {\rm eV/cm}^3}\ {\rm s}^{-1},
\end{equation}
where
\begin{equation}
w_0=w_{\rm B}+w_{\rm CMB}+w_{\rm opt}
\end{equation}
and (again following the estimates of Ref.~\cite{1995PhRvD..52.3265A}) where $w_{\rm CMB}\simeq0.25\ {\rm eV/cm}^3$  and $w_{\rm opt}\simeq0.5\ {\rm eV/cm}^3$ indicate the CMB and optical-IR radiation energy density, while $w_B\simeq0.6\ {\rm eV/cm}^3$ (for an average $B=5\mu$G) is the energy density associated to galactic magnetic fields. Altogether, we rather conservatively assume $p_2\simeq5.2\times 10^{-20}\ {\rm s}^{-1}$.

Under the assumption of a power law injection spectrum, the solution to eq.~(\ref{eq:diffusion}) can be readily calculated to be \cite{1995PhRvD..52.3265A}:
\begin{equation}\label{eq:master}
f(r,t,\gamma)=\frac{N_0\gamma^{-\alpha}}{\pi^{3/2}r^3}\left(1-p_2\ t\ \gamma\right)^{\alpha-2}\left(\frac{r}{r_{\rm dif}}\right)^3\exp\left(-\left(\frac{r}{r_{\rm dif}}\right)^2\right),
\end{equation}
with $\gamma<\gamma_{\rm cut}\equiv(p_2 t)^{-1}$, and $f=0$ otherwise, and where
\begin{equation}
r_{\rm dif}=2\sqrt{D(\gamma)\times t\times \frac{1-\left(1-\gamma/\gamma_{\rm cut}\right)^{1-\delta}}{\left(1-\delta\right)\gamma/\gamma_{\rm cut}}}.
\end{equation}
Assuming a pulsar burst-like injection at $t=T$, where $T$ is the age of the pulsar, the local spectrum of $e^\pm$ from pulsars is entirely fixed once the distance $Dist=r$ to the pulsar, its age and the normalization to the injected spectrum are given. THis approximation is well justified for mature pulsars, where $T$ is much larger than the trapping time of $e^\pm$ before they diffuse in the ISM, given the steep breaking index of dipolar emission as a function of time. The local $e^\pm$ flux will then simply be: 
\begin{equation}
J=c\times f(Dist,T,\gamma)/(4\pi).
\end{equation}
The first important feature of eq.~(\ref{eq:master}) above is that the pulsar spectra have a sharp cutoff (possibly smoothed out by deviations from the power-law behavior in the injected $e^\pm$ at high energy, expected in the very high energy regime, and by non-burst-like injections), and a normalization that depends only on the (energy dependent, unless $\delta=0$) factor $s=r/r_{\rm dif}$. 

As clear from the discussion above, the pulsar distance is therefore, clearly, a parameter of crucial importance in the estimate of the flux of the pulsar-injected electrons and positrons. For consistency, we refer to all pulsar distances as listed in the ATNF catalogue. Since, later on, we will further discuss the case of Geminga, and given that the estimate of the $e^\pm$ flux sensibly depends upon distance, we comment further on the value of 160 pc listed in the ATNF catalogue. This value refers to the result of optical measurements of the annual parallax of Geminga with data from Hubble Space Telescope observations \cite{gem1}, yielding a distance of $157^{+59}_{-34}$ pc. The range is compatible with the more recent and more conservative result reported in \cite{gem2}, $250^{+120}_{-62}$ pc. The large error-bars in the last measurement should serve as a warning to the reader as to the significant uncertainty in the measurement of pulsar distances, even in the case of well-known and studied objects like Geminga. Distances inferred from dispersion measures rely on the measurement of the mean electron number density at different places in the Galaxy, and are believed to be affected by errors as large as a factor 2 \cite{1986bhwd.book.....S}.

In addition to uncertainties in the {\em current} distance of pulsars, an important role in the estimate of the cosmic-ray electrons and positrons injected by mature pulsars is played by the pulsars' velocities. Pulsars have an average velocity at birth of around 250-300 km/s \cite{1997MNRAS.291..569H}, and a distribution of observed velocities today consistent with a mawellian distribution with a large width, of around 200 km/s. In addition, several outliers have been observed, with velocities even in excess of 1000 km/s. This means, for instance, that a pulsar with the current distance of Geminga, and with an age close to that estimated by its timing parameters, could have been almost anywhere in the sky at birth, and hence when it injected in the ISM the high energy $e^\pm$ population we are interested in here. For instance, Ref.~\cite{gem1} points out that if Geminga has a large enough transverse velocity ($\gtrsim700$ km/s), it could have traveled, since its birth, 250 pc and have been potentially located, at birth, at coordinates diametrically opposite to those observed today. The issue of pulsar proper velocities is clearly a very relevant one in the game of determining if a given pulsar is the origin of the observed lepton anomalies. This issue is particularly acute for nearby and relatively old pulsars, while is less severe for more distant pulsars. In this respect, the analysis we present below implicitly assumes, for nearby pulsars, a small transverse velocity. We show quantitatively the effect of pulsars' velocities on the pulsar age-versus-distance parameter space in fig.~\ref{fig:bestfit_distage}.

\begin{table}
\caption{\label{tab:diff} The three diffusion setups we consider in the present analysis. $D_1$ indicates the value of the diffusion coefficient at $E_{e^\pm}=1$ GeV, in units of cm${}^2/$s. }
\begin{ruledtabular}
\begin{tabular}{lcc}
 Scenario & $D_1$ & $\delta$\\ \hline
MAX & $1.8\times 10^{27}$ & 0.55\\
MED & $3.4\times 10^{27}$ & 0.7\\
MIN & $2.3\times 10^{28}$ & 0.46\\
\end{tabular}
\end{ruledtabular}
\end{table}

The last ingredient before starting our estimates is to define the diffusion coefficient. Since we only care about $E_{e^\pm}\gtrsim 10$ GeV, we neglect not only the effects of solar modulation, but also the energy-independent part of $D(\gamma)$ that kicks in at low energies, and assume $D(\gamma)\propto\gamma^\delta$, with the combination of normalizations and values of $\delta$ specified in tab.~\ref{tab:diff}, for the three diffusion setups MIN, MED and MAX. These three scenarios are such that simulations of various cosmic ray species abundances with those setups reproduce available cosmic-ray data, according to the analysis of Ref.~\cite{Delahaye:2007fr}.

Notice that the analytical solution of Eq.~(\ref{eq:master}) assumes spherical symmetry, an assumption which can clearly break down for the geometry of the diffusive region of the Galaxy, ordinarily assumed to be cylindrical, with a diffusive region of height $L$. Specifically, the three diffusion models of Tab.~\ref{tab:diff} correspond to half-thiknesses of the diffusion zone $L=1,\ 4$ and 15 kpc. Imposing free escape boundary conditions, namely that the cosmic ray density equals zero at the surfaces of the diffusive region (i.e. on the planes at $z=\pm L$) can be enforced exactly by resorting to a set of image charges similar to the one outlined in Ref.~\cite{Baltz:1998xv}. Specifically, we need to add an infinite series of image-pulsars identical to the one under consideration (that injects $e^\pm$ at a time $T$, with a given injection spectrum etc.) at $z_n=2Ln+(-1)^n z_0$, where $z_0$ is the distance of the original pulsar from the galactic plane, and adding all image-pulsars (for $-\infty<n<+\infty$) with a factor $(-1)^n$ \cite{Baltz:1998xv}. Assuming $z_0\sim0$, this amounts to simply account for the suppression factor ${\rm cyl}(\xi)$ given by
\begin{equation}
{\rm cyl}(\xi)=\sum_{n=-\infty}^{+\infty}\ \left(-1\right)^{n}\ \exp\left(-\xi^2 n^2\right),
\end{equation}
where $\xi$ defined as
\begin{equation}
\xi\equiv\frac{2L}{r_{\rm dif}(\gamma,T)}.
\end{equation}
In the general case $z_0\neq0$, we simply have
\begin{equation}
{\rm cyl}(\xi,z_0)=\sum_{n=-\infty}^{+\infty}\ \left(-1\right)^{n}\ \exp\left(-\xi^2 n^2-2(-1)^n\xi n \frac{z_0}{L}\right).
\end{equation}
The Reader should bear in ming that the method outlined above neglects the radial boundary condition usually considered in diffusion models. This is well justified if the sources under consideration are closer to the observer than to the radial boundary of the diffusive region \cite{Baltz:1998xv}. This is a good approximation with the galactic sources under consideration here (see e.g. fig.~10.5 of \cite{1986bhwd.book.....S}). Interestingly, ${\rm cyl}(\xi)$ doesn't depend on the pulsar distance, but only on the ratio of the thickness of the diffusion zone over the effective diffusion radius $r_{\rm dif}$. Quantitatively, this statement holds as long as the pulsar distance is small compared to the neglected radial boundary condition. 

\begin{figure}
\includegraphics[width=9.cm,clip]{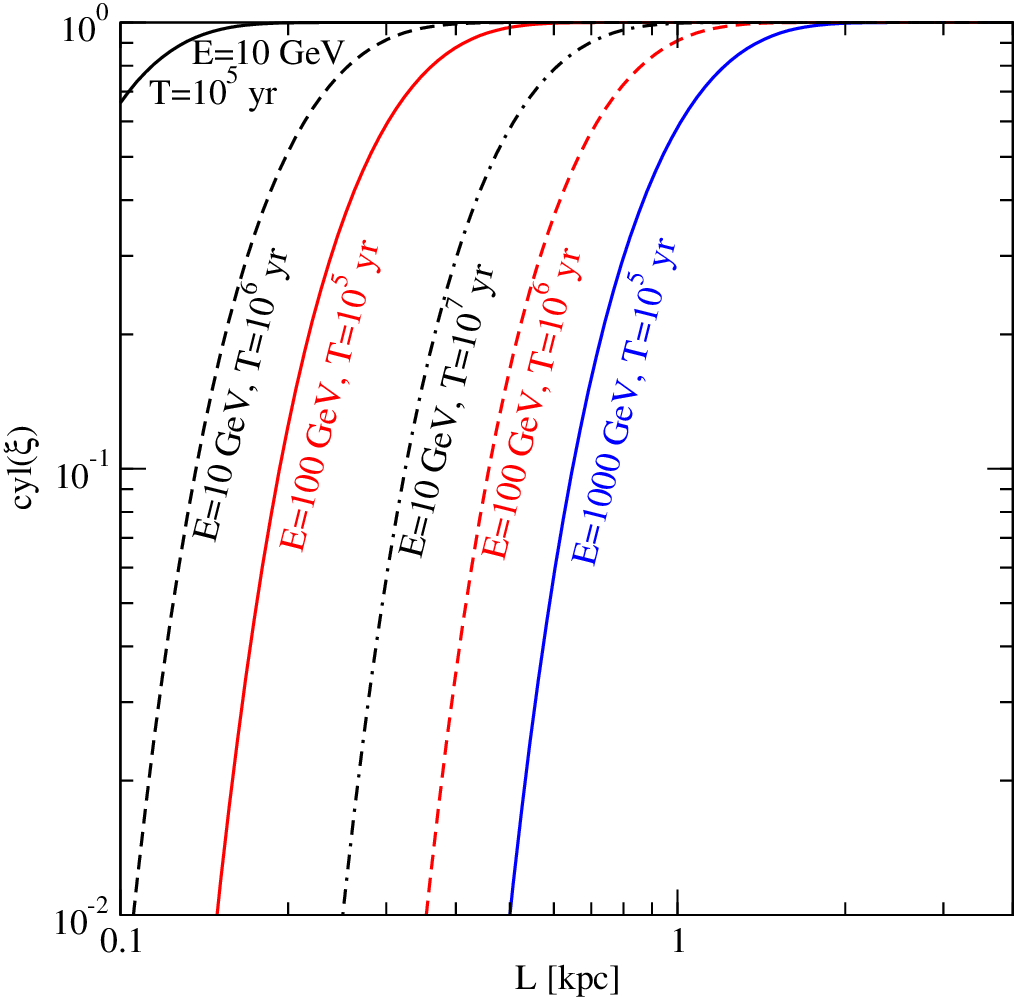}
\caption{\label{fig:spherical} The effect of electron-positron escape from the Galaxy (defined by the suppression fraction ${\rm cyl}(\xi)$), as a function of the thickness of the galactic diffusion region $L$, for selected pairs of values for the injection time and energy.}
\end{figure}

Fig.~\ref{fig:spherical} illustrates the effect of deviations from the spherically symmetric solution as a function of $\xi$, as well as for selected pairs of pulsar ages and $e^\pm$ energies, as a function of $L$ (we employ the values of the MED parameters for this plot, but let $L$ vary). As a rule, the breakdown of spherical symmetry must be taken into account when $\xi\lesssim 0.5$, and a source becomes essentially irrelevant if $\xi\lesssim0.2$. Effects of order 10\% are expected for diffusion regions with a thickness between 1 and 2 kpc, and for the high-energy end ($E_{e^\pm}\sim10$ GeV) of the spectrum of young pulsars, or for the low energy end ($E_{e^\pm}\sim10$ GeV) of mature pulsars. For the following discussion, escape of pulsar-injected $e^\pm$ is therefore typically not relevant.

To summarize, once a theoretical model is chosen, the contribution of pulsars to the $e^\pm$ flux depends on three observational inputs: the pulsar age (as determined from timing measurements), its distance and its spin-down power (which can be connected, assuming a particular model such as one of those outlined here above, to the total $e^\pm$ energy output injected in the ISM). To compute the resulting flux at Earth, one further needs to specify a diffusion setup: we restrict this choice to the three possibilities of tab.~\ref{tab:diff}. We are now ready to calculate the contribution of some of the most likely sources of primary energetic positrons (and electrons): nearby bright pulsars.

\section{Contributions from selected nearby pulsars}\label{sec:nearby}

\begin{table*}
\caption{\label{tab:nearby} Data for a few selected nearby pulsars and SNR's. $E_{\rm out}$ is the energy output in $e^\pm$ pairs in units of $10^{48}$ erg (for the ST model column we assumed $f_{e^\pm}=3\%$). The energy output for the SNR Loop I and Cygnus Loop are not estimated within the ST model, but via estimates of the total SNR output. The $f_{e^\pm}$ column indicates the $e^\pm$ output fraction used to compute the fluxes shown in fig.~\ref{fig:spectrum_nearby} and \ref{fig:ratio_nearby} assuming the ST model.}
\begin{ruledtabular}
{\scriptsize
\begin{tabular}{lccccccccc}
 Name & Distance [kpc] & Age [yr] & $\dot E$ [ergs/s] & $E_{\rm out}$ [ST] & $E_{\rm out}$ [CCY] & $E_{\rm out}$ [HR] &$E_{\rm out}$ [ZC] & $f_{e^\pm}$ & $g$\\ \hline
Geminga [J0633+1746]& 0.16 & $3.42\times 10^5$ & $3.2\times10^{34}$ & 0.360 & 0.344 & 0.013 & 0.053 & 0.005 & 0.70\\
Monogem [B0656+14] & 0.29 & $1.11\times 10^5$ & $3.8\times10^{34}$ & 0.044 & 0.133 & 0.006 & 0.020 & 0.020 & 0.70\\
Vela [B0833-45] & 0.29 & $1.13\times 10^4$ & $6.9\times10^{36}$ & 0.084 & 0.456 & 0.006 & 0.372 & 0.0015 & 0.14\\
B0355+54 & 1.10 & $5.64\times 10^5$ & $4.5\times10^{34}$ & 1.366 & 0.677 & 0.022 & 0.121 & 0.2 & 0.61\\ \hline
Loop I [SNR] & 0.17 & $2\times 10^5$ & & {\em 0.3} &  &  &  & 0.006 & \\
Cygnus Loop [SNR] & 0.44 & $2\times 10^4$ & & {\em 0.03} &  &  &  & 0.01 & \\
\end{tabular}
}
\end{ruledtabular}
\end{table*}

\begin{figure*}
\includegraphics[width=14.cm,clip]{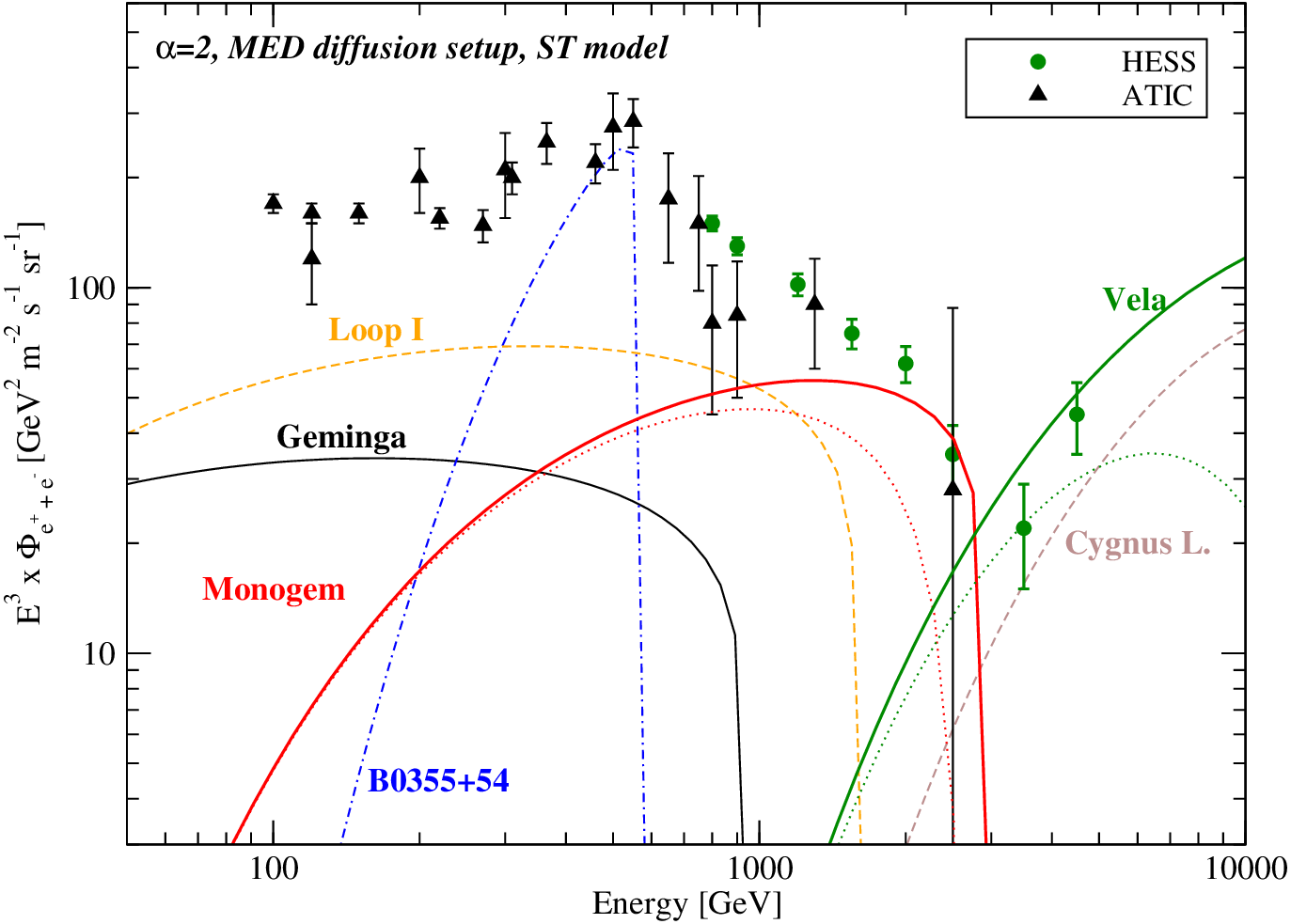}
\caption{\label{fig:spectrum_nearby}The spectrum of selected nearby pulsars and SNR's (for the parameters employed to calculate the fluxes see tab.~\ref{tab:nearby}). We assume an $e^\pm$ injection spectral index $\alpha=2$, and a median diffusion setup (MED). The dotted lines correspond to injection spectra featuring an exponential cutoff at $E_{e^\pm}=10$ TeV.}
\end{figure*}
\begin{figure}
\includegraphics[width=10.cm,clip]{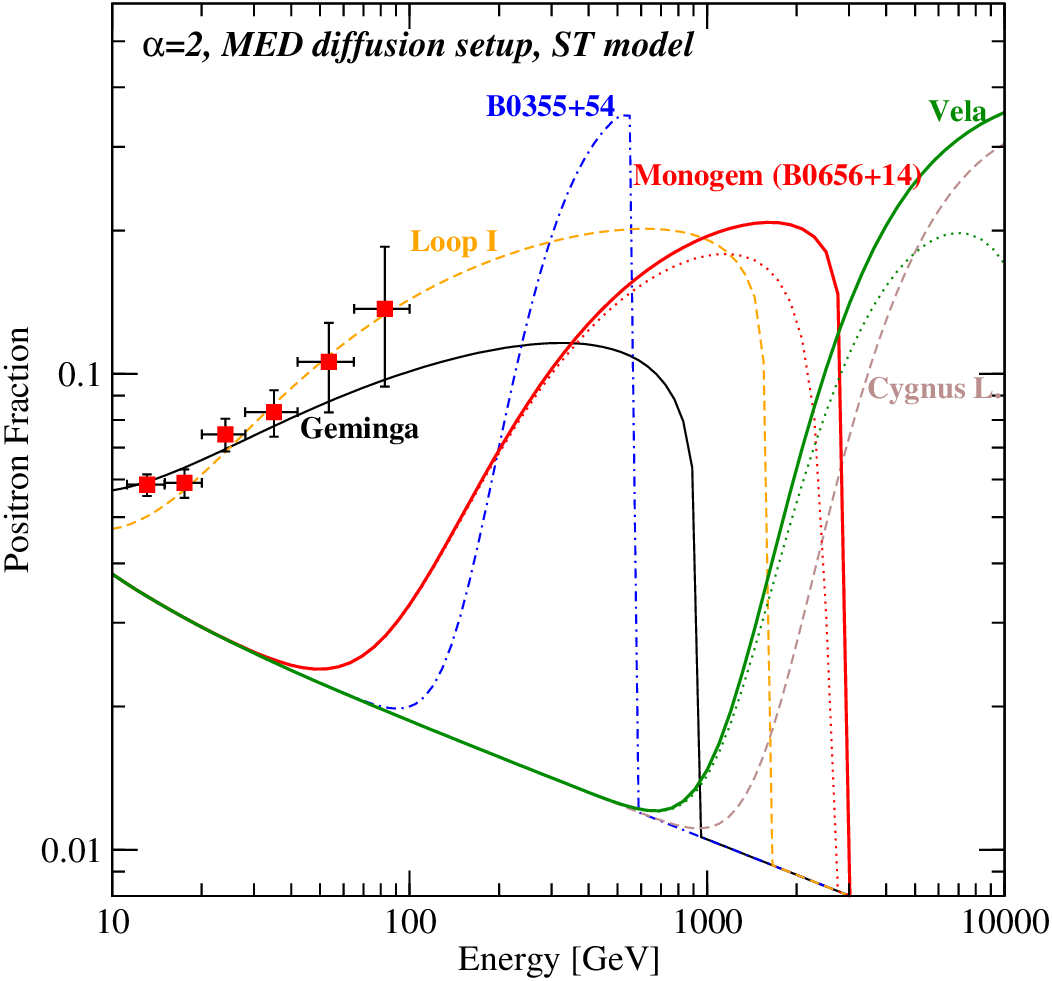}
\caption{\label{fig:ratio_nearby}The positron fraction for the same sources as in fig.~\ref{fig:spectrum_nearby} and tab.~\ref{tab:nearby}.}
\end{figure}

In this section we apply the setup outlined above to a few objects that, naively, can contribute to the $e^\pm$ flux needed to account for the PAMELA and ATIC data. Specifically, we consider the Geminga [J0633+1746], the Monogem [B06546+14] and the Vela [B0833-45] pulsars: all these pulsars lie within 300 pc, and although Vela might be too young to contribute significantly to the $e^\pm$ flux in the energy range of interest, and the produced $e^\pm$ might not have diffused in the ISM yet, all these pulsars have a very large spin-down energy loss rate. In addition, we consider the radio pulsar PSR B0355+54, a middle aged ($T\sim0.56$ Myr) pulsar with a fairly large spin-down energy loss rate and a relatively small distance. We are lead to consider this particular pulsar to explain the ATIC ``bump'' feature, i.e. the reported excess over the expected $e^\pm$ flux. Naively, a pulsar that contributes to the bump, assuming a relatively hard injection spectrum, should have an age
\begin{eqnarray}
 &T_{\rm ATIC}\sim m_e/\left(p_2\times E_{\rm bump}\right)\approx 5-6\times10^5\ {\rm yr}&\\ 
\noindent &{\rm for}\quad E_{\rm bump}\sim500-600\ {\rm GeV}.&
\end{eqnarray}
In addition, the candidate pulsar should be distant enough for the resulting spectrum to be peaked enough around $E_{\rm bump}$ to avoid excessive contributions at lower energies, and bright enough to produce the right level of $e^\pm$. In short, PSR B0355+54 appears to be a natural candidate fulfilling all of these requirements, which is why we consider it here (notice that in what follows we consider the possibility of multiple pulsars explaining the ATIC feature as well). Finally, we also include the two SNR's Loop I and Cygnus Loop, following the analysis of Ref.~\cite{2004ApJ...601..340K}. As mentioned in the introduction, radioactive decay, or even a faint and yet undetected pulsar associated to these SNR could inject enough positrons for the two SNR's to contribute significantly to the positron fraction as well.

We outline the set of pulsars for which we shall estimate the positron fraction and the total $e^\pm$ spectrum in tab.~\ref{tab:nearby}. The table indicates the name, distance, age and the spin-down power of the various pulsars. The following four columns specify the estimated total $e^\pm$ energy output according to the four models described in sec.~\ref{sec:models}, with our standard assumptions for the relevant model parameters. Finally, the $f_{e^\pm}$ column indicates the actual value for the efficiency factor in the $e^\pm$ output within the ST model. In all cases we adopt values around the estimate quoted in Ref.~\cite{Hooper:2008kg} and in Ref.~\cite{Malyshev:2009tw}. In one case, for PSR B0355+54 we adopt a factor 6-7 larger efficiency output. This should be contrasted with the variance in the model estimates for the total $e^\pm$ output of pulsars, but it certainly indicates an unusually large efficiency for this single pulsar to produce enough $e^\pm$ to account for the ATIC bump.

Fig.~\ref{fig:spectrum_nearby} shows the resulting $e^\pm$ differential energy spectrum (times energy to the third power) for the six sources listed in tab.~\ref{tab:nearby}, as well as the ATIC \cite{2008Natur.456..362C} and the HESS data \cite{Collaboration:2008aa}. The following fig.~\ref{fig:ratio_nearby} shows the estimated positron fraction, with the PAMELA data \cite{Adriani:2008zr}. We assume a median spectral index $\alpha=2$ for all sources, and employ the MED diffusion setup and the ST $e^\pm$ output model, with the values of $f_{e^\pm}$ specified in tab.~\ref{tab:nearby}. For illustrative purposes, we also show, with dotted thin lines the effect of a cutoff at $E_{\rm cut}=10$ TeV for the Monogem and Vela pulsars.

The figures show that both the nearby Geminga pulsar and the Loop I SNR, both at distances around 160-170 pc, and both with ages $T\sim2-3\times 10^5$, can very naturally be the dominant positron sources to explain the PAMELA data, with an output in $e^\pm$ less than a percent of the pulsar rotational energy, in line with, or less than, previous estimates of that efficiency factor. The resulting contribution to the total $e^\pm$ at larger energies is subdominant. Young, powerful and relatively nearby sources like Vela and the Cygnus Loop likely contribute to the highest energy part of the total $e^\pm$ differential spectrum. Fig.~\ref{fig:spectrum_nearby} seems even to suggest that the highest energy H.E.S.S. bin could even actually reflect the contribution from these bright local sources. The contribution of these sources to the positron fraction kicks in above the TeV scale, and is therefore irrelevant for the positrons detected by PAMELA. The Monogem SNR pulsar might plausibly contribute to the higher energy bins reported by PAMELA, and could be dominating the $e^\pm$ flux between 1 and 3 TeV, even with a $e^\pm$ efficiency factor of $f_{e^\pm}\simeq1.5\%$. Finally, the ATIC bump seems to fit very naturally with the expected $e^\pm$ flux from the PSR B0355+54, although this requires a very efficient, but still possible in principle, output of $e^\pm$ ($f_{e^\pm}\sim20\%$). The positrons produced by PSR B0355+54 would give a sizable and steeply rising contribution to the positron fraction between 200 GeV and 600 GeV, a range that might eventually be explored by PAMELA \cite{Adriani:2008zr}.

In short, in this section we showed that selected, known nearby $e^\pm$ sources, with a reasonable $e^\pm$ energy output, very naturally account for both the PAMELA and the ATIC data. None of these sources appears to be suitable to explain both experimental results at once. In the next section we address the question of whether one single $e^\pm$ source (combing the entire pulsar parameter space) can explain at once the PAMELA and the ATIC data.
 
\section{Minimalism: Single Source}\label{sec:single}

\begin{figure*}
\includegraphics[width=17.cm,clip]{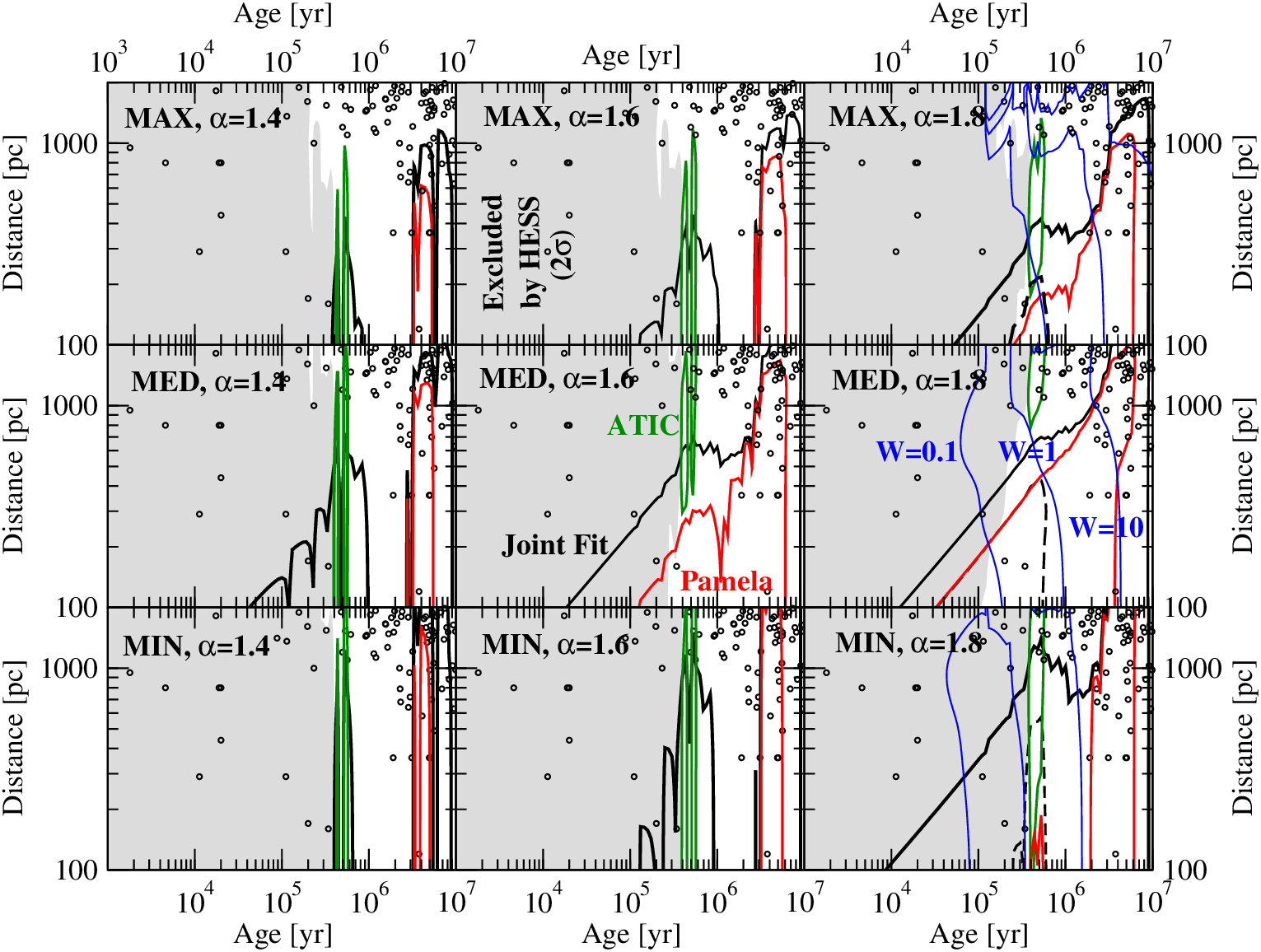}
\caption{\label{fig:single_hard}Likelihood contours for a single pulsar-like source normalized to give the best joint fit to the PAMELA and to the ATIC data, in the age versus distance plane. The best fit regions ($\chi^2/{\rm d.o.f.}<3$) are below the black curves. The regions within the red and green lines have a $\chi^2/{\rm d.o.f.}<1$ for the fit to the ATIC and to the PAMELA data alone, respectively. The grey region is excluded by the HESS data \cite{Collaboration:2008aa}. The blue contours indicate the inferred $e^\pm$ energy output, in units of $10^{48}$ erg. The black circles indicate the age and distance of known pulsars in the ATNF catalogue. In the three upper panels we assume the MAX diffusion setup, in the central ones the MED setup and in the three lower panels the MIN setup. The three panels to the left have $e^\pm$ injection spectral index $\alpha=1.4$, those in the middle 1.6 and those to the right 1.8.  }
\end{figure*}

\begin{figure*}
\includegraphics[width=17.cm,clip]{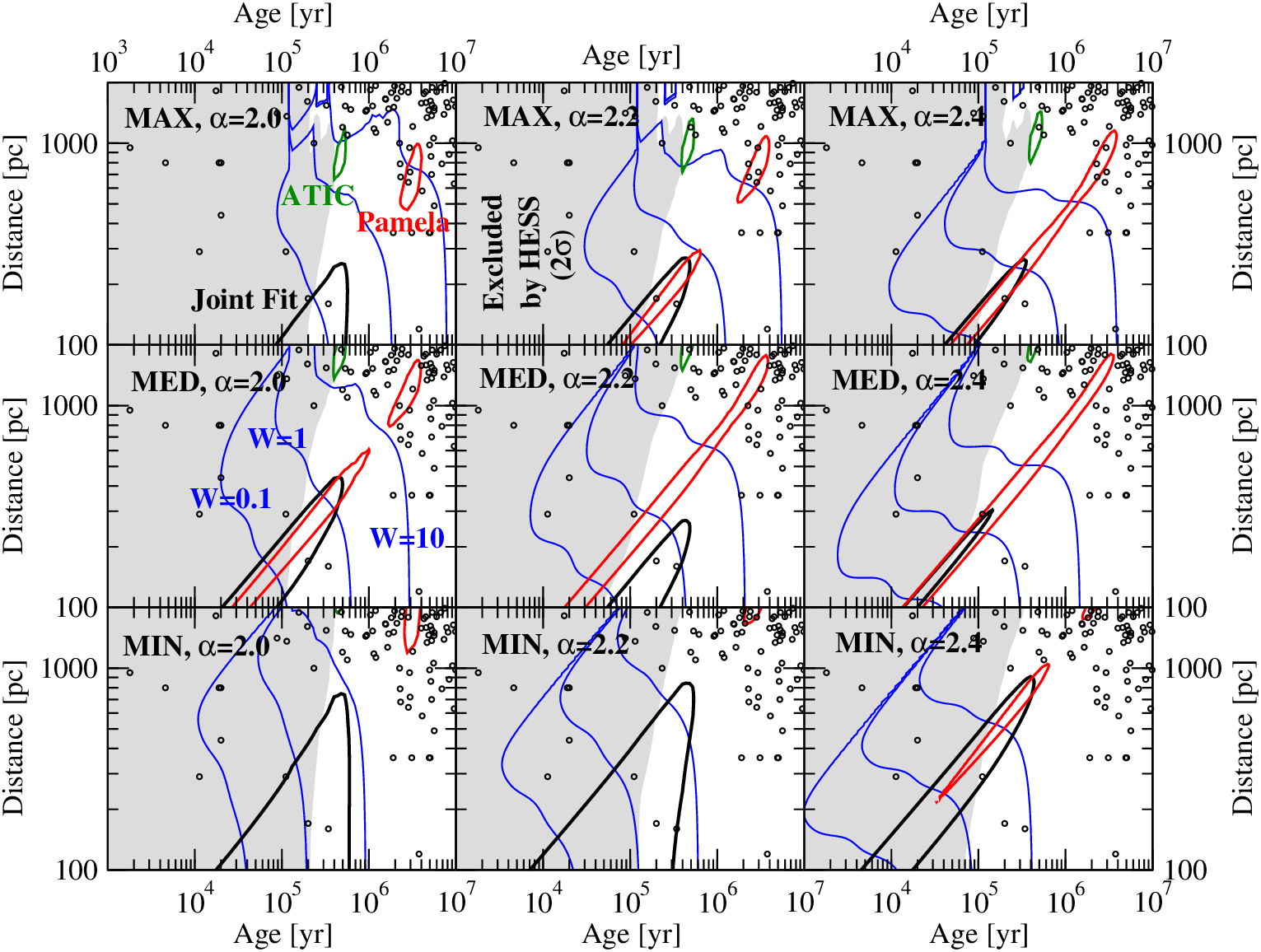}
\caption{\label{fig:single_soft}As in fig.~\ref{fig:single_hard}, for softer $e^\pm$ injection spectral indexes (2, 2.2 and 2.4, from left to right).}
\end{figure*}

In fig.~\ref{fig:single_hard} and \ref{fig:single_soft} we explore the possibility that the ATIC and PAMELA data originate from one single source. We fix the injection spectrum of $e^\pm$ at the source to $\alpha=$1.4, 1.6, 1.8 (fig.~\ref{fig:single_hard}) and to $\alpha=$2.0, 2.2, 2,4 (fig.~\ref{fig:single_soft}). For each value of $\alpha$, in the top, middle and bottom panels we adopt the MAX, MED and MIN diffusion setups, respectively. On each panel, we normalize the source $e^\pm$ energy output to get the best fit (i.e. the lowest $\chi^2$ per degree of freedom (d.o.f.)) to both the ATIC and the PAMELA data. The regions shaded in grey have fluxes, in at least one of the H.E.S.S. data bins, in excess of the measurement by 2-$\sigma$, and we therefore consider these regions excluded. The black contours indicate a global $\chi^2$ per degree of freedom less or equal than 3. The best fit regions are therefore inside the black contours. Notice that we verified that including other data sets on the high-energy flux of electrons and positrons (e.g. the recent PPB-BETS measurement \cite{Torii:2008xu}) does not visibly affect the contours corresponding to the ATIC data alone: this simply depends on the much better statistics that the ATIC measurements have with respect to previous data.

The regions inside the green and the red contours have a $\chi^2$ per degree of freedom less or equal than one for the individual ATIC and PAMELA data-sets, respectively. This means that, by normalizing the source to the best fit to {\em both} data sets, the regions giving the best fits to the individual ATIC and PAMELA data are those within the green and red contours. 

Finally, we show contours of constant $e^\pm$ energy output, which we indicate as $W=E_{\rm out}/10^{48}\ {\rm erg}$. Roughly, a reasonable range for $W$ is between 0.1 and 10. We also indicate with black circles the actual location of the ATNF pulsars in the age versus distance plane. We, again, stress that those points locating pulsars in the age-distance parameter space are subject to (i) uncertainties in the determination of the current distance of the pulsars, and on the determination of the age (what is shown is the characteristic age estimated from time measurements), and (ii) a systematic uncertainty from the individual pulsar velocities (this effect is particularly relevant for nearby and mature pulsars, hence for those points in the lower-right corner of each panel). 

We also remark that the contours we show in fig.~\ref{fig:single_hard} and \ref{fig:single_soft} would be naturally affected by changing the astrophysical background model for the diffuse cosmic-ray electron and positron component. Specifically, we concentrate on what would happen if one changed the normalization and the spectral index of the diffuse galactic high-energy electron-positron spectrum. We verified that a shift in the local normalization almost only affects the contours by shifting the contours of iso-level for the variable $W$, but does not affect significantly the shape of the regions preferred by an explanation to the PAMELA and ATIC anomalies. A tilt in the diffuse background does affect the shape of the contours. Remarkably, though, this effect amounts to shifting the contours towards harder or softer values for $\alpha$, that compensate the variation in the background spectrum. In essence, we find that changing the background spectrum corresponds to variations in the location and shape of the contours favored by the lepton anomalies comparable to what shown, from left to right, in fig.~\ref{fig:single_hard} and \ref{fig:single_soft}.

 With a hard spectrum, fig.~\ref{fig:single_hard}, clearly a single source interpretation is disfavored: the ATIC data strongly prefer a pulsar with an age around 0.4 to 0.6 Myr, while PAMELA favors an older source with an age in the 3-6 Myr range\footnote{Notice that the precise value of these age ranges depends critically on the assumed $e^\pm$ quadratic energy loss coefficient $p_2$.}. As the spectrum gets softer, the PAMELA data start to be compatible with a broader combination of ages and distances. For $\alpha=1.8$, the Geminga pulsar appears to be roughly at the intersection of the PAMELA and ATIC $\chi^2/{\rm d.o.f.}=1$ regions.
 
 Moving to softer injection spectra (fig.~\ref{fig:single_soft}), the ATIC data favor a very distinct region in the age-distance plane, again with ages around 0.4 to 0.6 Myr, and distance ranges which depend on the diffusion setup and on the injection spectral index. PAMELA points toward a relatively wide range of very correlated values for age and distance. Several pulsar candidates fall within both the ATIC and the PAMELA $\chi^2/{\rm d.o.f.}=1$ candidates, but again, the overall information we extract is that the two regions are fairly distant. The ATIC and PAMELA data seem to most naturally require two distinct sources.
 
 As a last comment, looking at the blue lines with contours of output power, the regions favored by ATIC tend to require very large $e^\pm$ energy outputs, especially with softer injection spectra, pointing to the fact that {\em the pulsar scenario is compatible, but does not favor, for natural values of the parameters, a peaked spectrum such as what reported by ATIC}. 
 
 Given that a single source interpretation gives only a partly satisfactory fit to the data, we now move on to explore the region of pulsar parameter space which are, separately, favored by the PAMELA (sec.~\ref{sec:pamela}) and by the ATIC (sec.~\ref{sec:atic}) data.

\section{Pamela and Occam}\label{sec:pamela}

\begin{figure}[!h]
\mbox{\includegraphics[width=8.cm,clip]{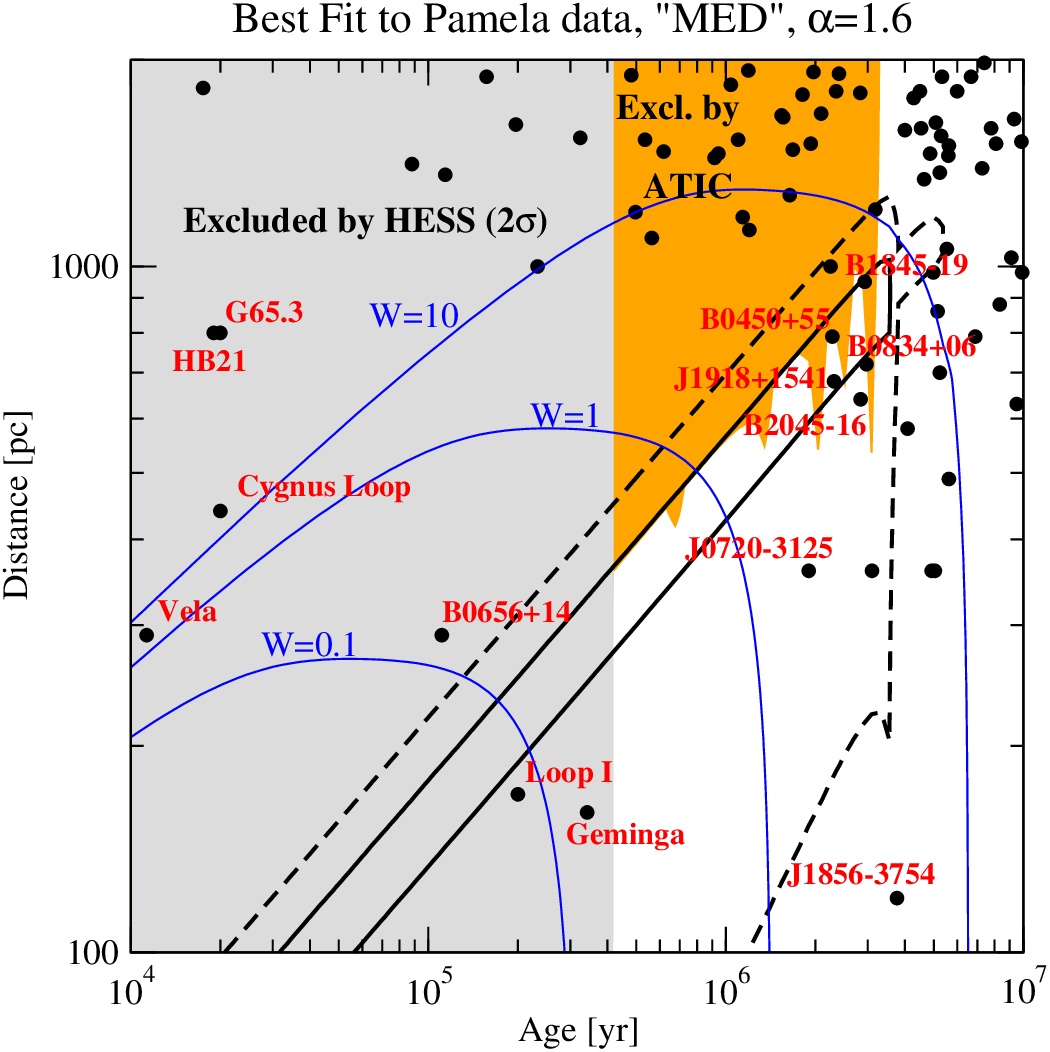}\quad\includegraphics[width=8.cm,clip]{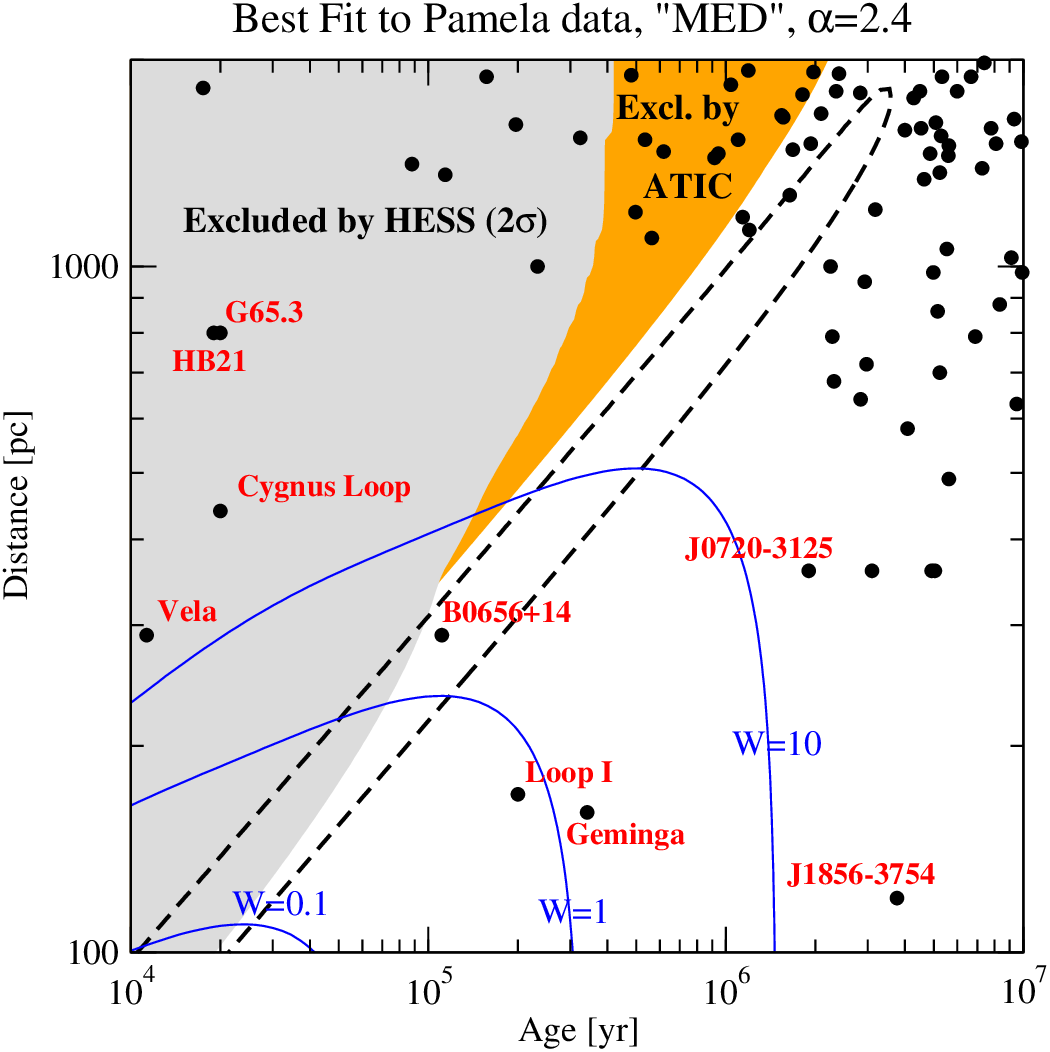}}\\[0.5cm]
\mbox{\includegraphics[width=8.cm,clip]{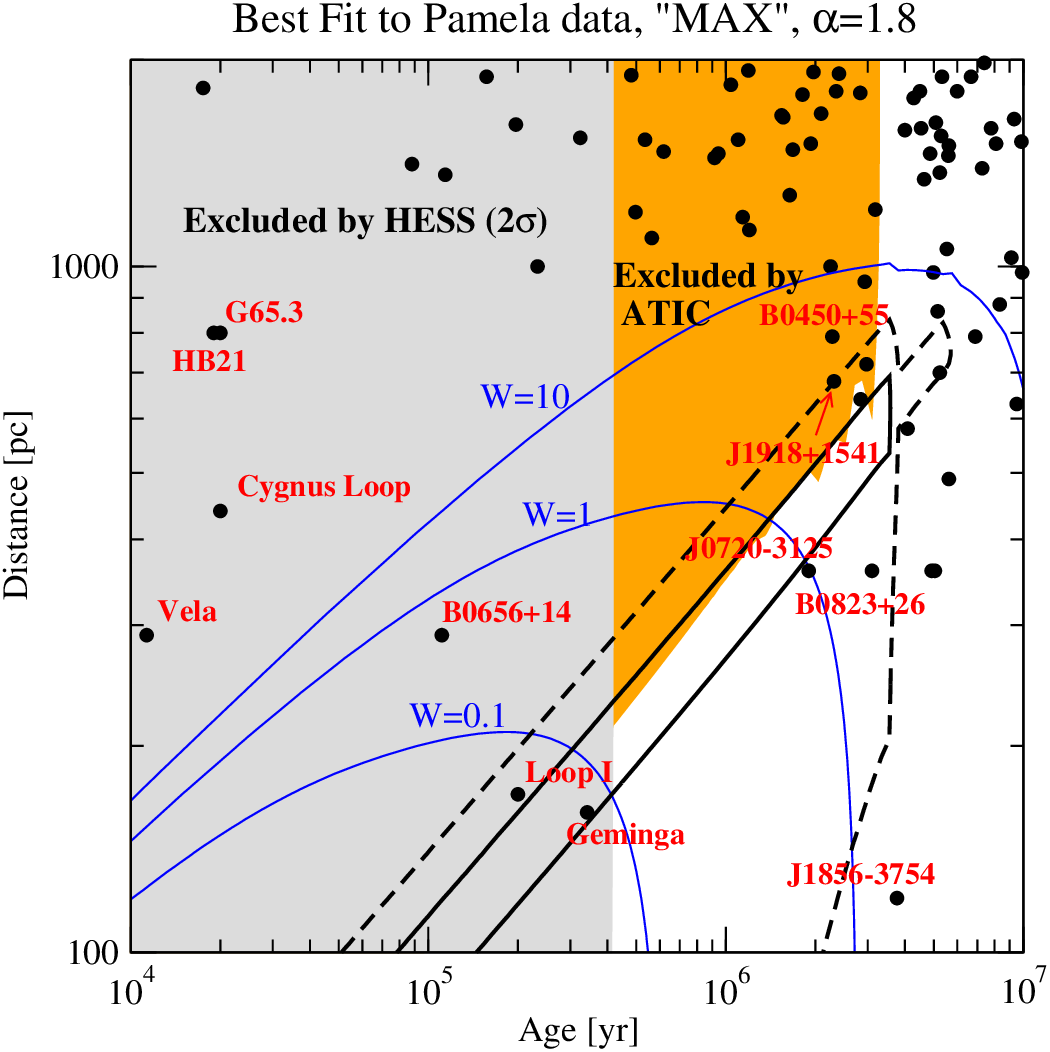}\quad\includegraphics[width=8.cm,clip]{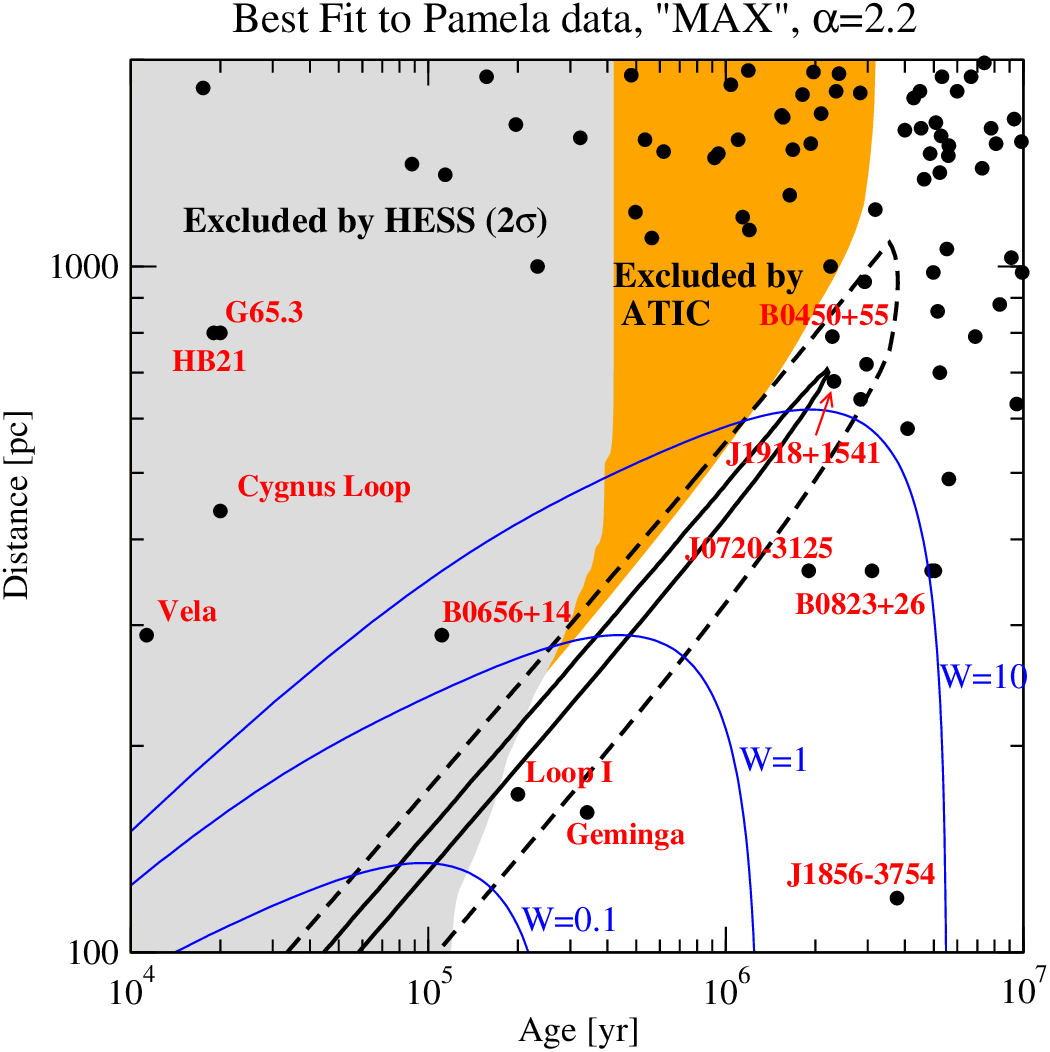}}
\caption{\label{fig:pamela}Regions providing the best fit, with a single source, to the PAMELA data \cite{Adriani:2008zr}. The grey region is ruled out by HESS, and the orange region is in conflict with the ATIC data. As in fig.~\ref{fig:single_hard}, we indicate (and sometimes label) the age and distance of known pulsars, as well as curves of constant inferred $e^\pm$ energy output, in units of $10^{48}$ erg. The region of parameter space inside the solid black lines has a $\chi^2/{\rm d.o.f.}<0.5$, while inside the black dashed region $\chi^2/{\rm d.o.f.}<1$.}
\end{figure}

The four panels of fig.~\ref{fig:pamela} outline, for various pairs of injection spectral indexes and diffusion setups, the region of the pulsar age versus distance parameter space favored by PAMELA and compatible with the ATIC and with the H.E.S.S. data.

The two panels at the top assume a MED diffusion setup with $e^\pm$ injection spectral index $\alpha=1.6$ and 2.4, while the two panels at the bottom employ a MAX diffusion setup, with $\alpha=1.8$ and 2.2. As in the figures above, the grey regions are in conflict with the H.E.S.S. data, and the blue lines indicate the source energy output in $e^\pm$, in units of $10^{48}$ erg. In addition, we shade in orange the regions disfavored by the ATIC measurements, and label some of the pulsars which might be relevant to explain the PAMELA signal. The dashed contour outlines the region inside which $\chi^2/{\rm d.o.f}<1$, while inside the black solid contours  $\chi^2/{\rm d.o.f}<0.5$.

With a hard $\alpha\sim1.4...1.6$ spectrum, the PAMELA data point to nearby pulsars (distance less than 1 kpc), with an age (correlated to the distance) which the H.E.S.S. data constrain to be larger than $\approx4\times 10^5$ yr, and that the PAMELA data force to be less than $\approx4\times 10^6$ yr. Consistency with the ATIC data, as well as the values for the energy output, point towards a multiple source contribution as the most likely situation. The first four entries of tab.~\ref{tab:combination} list a few plausible pulsars that can contribute to the PAMELA signal, assuming a MAX diffusion setup. The list includes PSR J1918+1541, B0450+55, B0834+06 and B1845-19. 

Softer spectra generically worsen the quality of the fit to the PAMELA data (of course, not all pulsars will have the same spectral index, but softer spectra are generically unlikely to add up to the positrons measured by PAMELA with the correct spectral shape). With the MED diffusion setup, the Monogem pulsar, with a reasonable $e^\pm$ energy output close to $\sim10^{48}$ erg, gives an acceptable fit to PAMELA. With a MAX diffusion setup, it instead seems more plausible that a series of pulsars with lifetimes of a few Myr and distances between 0.7 and 1 kpc are the most relevant contributors to the PAMELA signal.

In any case, we showed in this section that no matter which injection spectrum is assumed, and which diffusion setup is employed, known pulsars naturally predict a positron fraction close to what is reported by PAMELA, with very reasonable electron-positron energy outputs. 

\section{ATIC and Occam}\label{sec:atic}

\begin{figure*}
\mbox{\includegraphics[width=8.cm,clip]{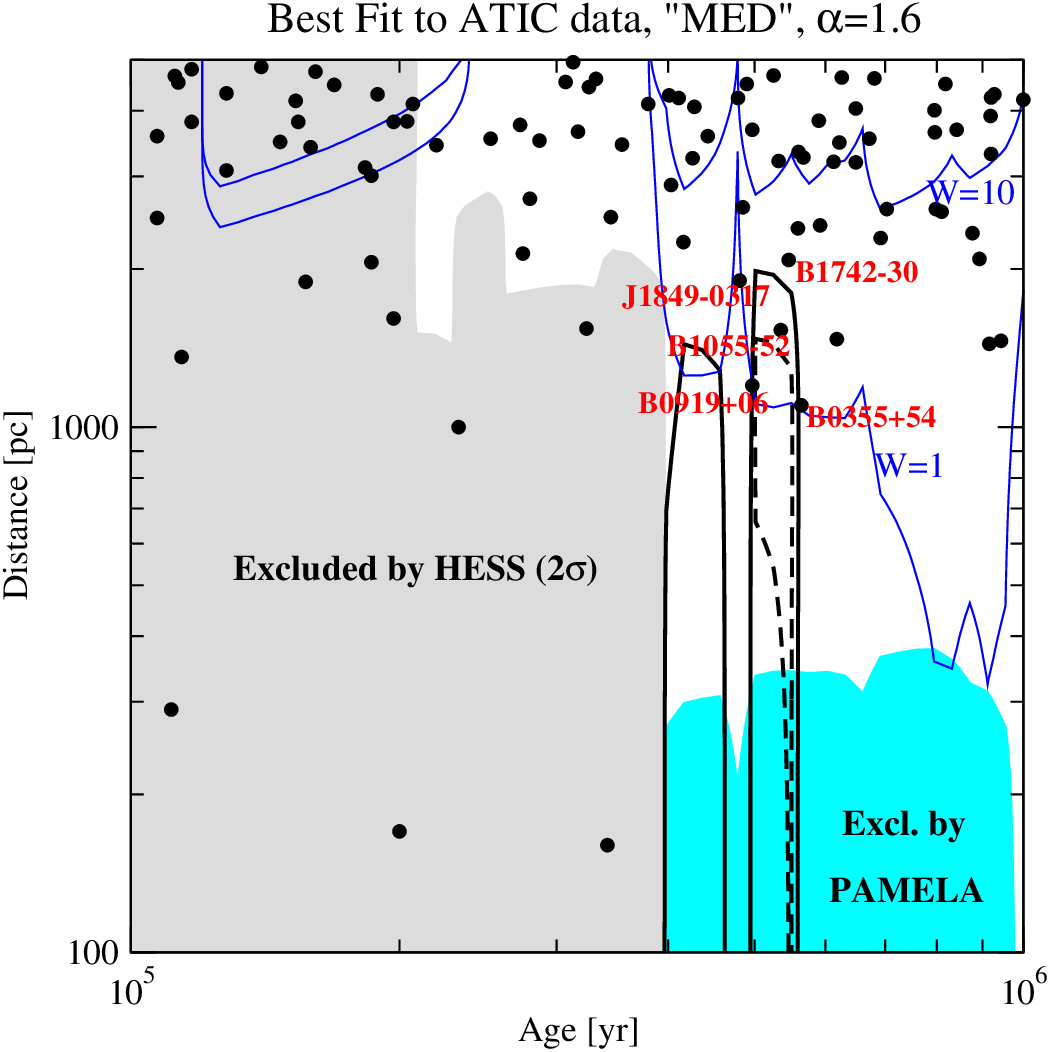}\quad\includegraphics[width=8.cm,clip]{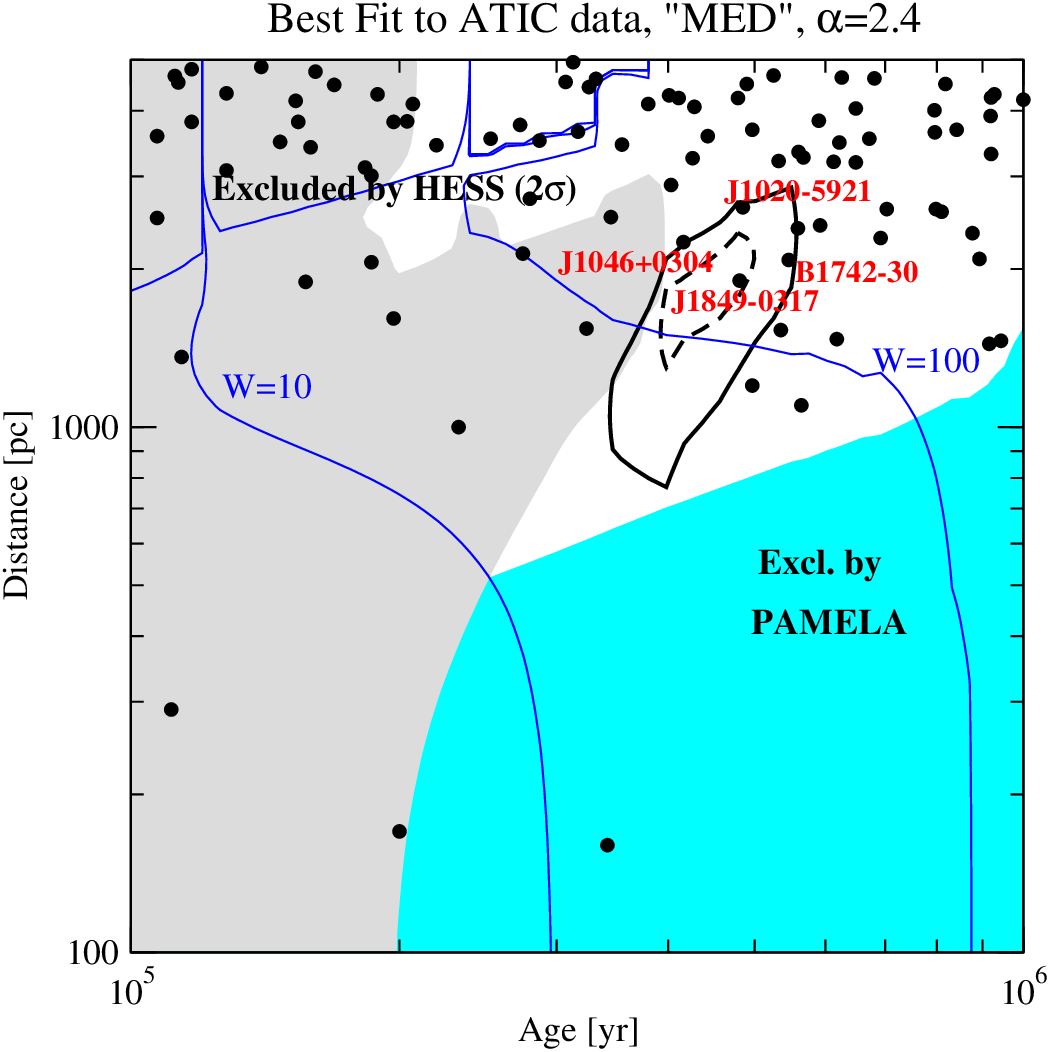}}\\[0.5cm]
\mbox{\includegraphics[width=8.cm,clip]{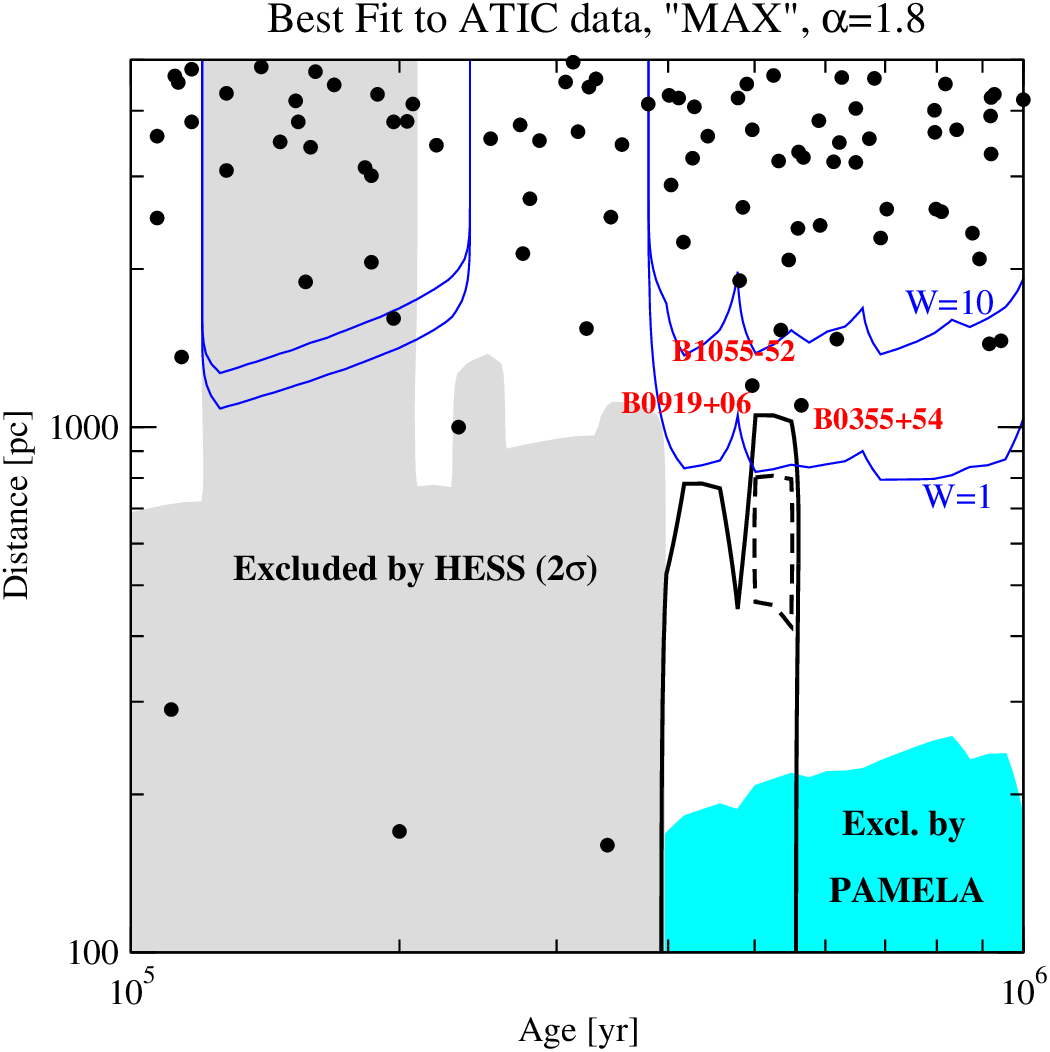}\quad\includegraphics[width=8.cm,clip]{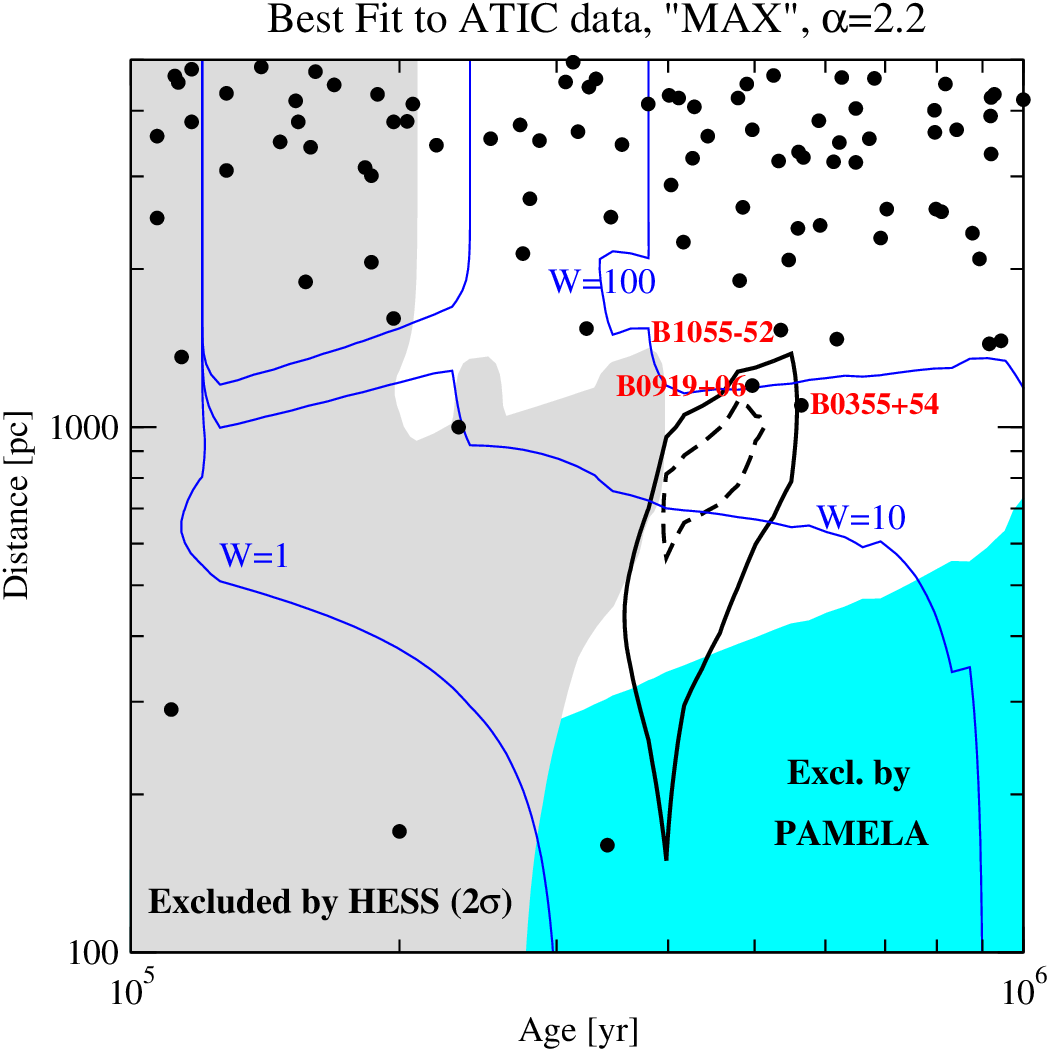}}
\caption{\label{fig:atic}Regions providing the best fit, with a single source, to the ATIC ``bump'' feature \cite{2008Natur.456..362C} (we define the bump feature as all ATIC data with $E_{e^\pm}>300$ GeV). The grey region is ruled out by HESS, and the cyan region is in conflict with the PAMELA data. As in fig.~\ref{fig:single_hard}, we indicate (and sometimes label) the age and distance of known pulsars, as well as curves of constant inferred $e^\pm$ energy output, in units of $10^{48}$ erg. The region of parameter space inside the solid black lines has a $\chi^2/{\rm d.o.f.}<0.5$, while inside the black dashed region $\chi^2/{\rm d.o.f.}<1$.}
\end{figure*}

Fig.~\ref{fig:atic} shows the portion of the pulsar age-distance parameter space favored by the ATIC data. The notation is as in the previous figure \ref{fig:pamela}, but now the regions shaded in cyan are disfavored by PAMELA. Independently of the injection spectrum, the inverse Compton plus Synchrotron cutoff, as mentioned above, dictates an age between 0.4 and 0.6 Myr. In addition, while distance matters less for a hard spectral index (as long as the pulsar $e^\pm$ output is large enough), for softer spectral indexes the ATIC data prefer distances of the order of 1 kpc.

An important message we read off of fig.~\ref{fig:atic} is that unless the injection spectrum is very hard (left panels), existing pulsars must be extremely powerful to account for the ATIC data: in the right panels the pulsars that lie within the $\chi^2/{\rm d.o.f}<1$ regions typically need normalizations $W\sim100$, i.e. an $e^\pm$ energy output of $E_{\rm out}\sim10^{50}$ erg, which, while possible in principle in some of the models we considered above (see fig.~\ref{fig:output}) seems unlikely for the standard energy budget of supernova explosions.

In summary, while the PAMELA data can be regarded as a natural {\em post}-diction of the pulsar scenario, the ATIC data find a possible explanation in this scenario as well, although energy output considerations indicate that the signal reported by ATIC would not be the most obvious to be expected from known pulsars.

\section{One Single, Nearby Source versus a Combination of Pulsars}\label{sec:onemulti}

\begin{table*}
\caption{\label{tab:combination} Possible combinations of multiple pulsars contributing to explain the PAMELA and the ATIC data. P/A refers to whether the pulsar dominantly contributes to the PAMELA or to the ATIC signal. $E_{\rm out}$ is the energy output in $e^\pm$ pairs in units of $10^{48}$ erg.}
\begin{ruledtabular}
\begin{tabular}{lccccccc}
 Name & P/A &Distance [kpc] & Age [yr] & $E_{\rm out}$ [ST] & $E_{\rm out}$ [CCY] & $E_{\rm out}$ [HR] &$E_{\rm out}$ [ZC] \\ \hline
J1918+1541 & P & 0.68 & $2.31\times 10^6$ & 0.99 & 0.33 & 0.023 & 0.022\\
B0450+55 & P & 0.79 & $2.28\times 10^6$ & 1.16& 0.37& 0.025& 0.025\\
B0834+06 & P & 0.72 & $2.97\times 10^6$ & 0.11& 0.07& 0.011& 0.001\\
B1845-19 & P & 0.95 & $2.93\times 10^6$ & 0.01& 0.015&0.005 &0.0002 \\ \hline
B0919+06 & A & 1.20 & $4.97\times 10^5$ & 0.158& 0.178& 0.010& 0.016\\
B0355+54 & A & 1.10 & $5.64\times 10^5$ & 1.366 & 0.677 & 0.022 & 0.121\\
B1055-52 & A & 1.53 & $5.35\times 10^5$ & 0.82& 0.49&0.017 &0.075 \\
J1849-0317 & A & 1.90 & $4.81\times 10^5$ & 0.06& 0.10& 0.007& 0.007\\
B1742-30 & A & 2.08 & $5.46\times 10^5$ & 0.24& 0.22& 0.012& 0.022\\
\end{tabular}
\end{ruledtabular}
\end{table*}

In this section we consider the coherent superposition of a few pulsars that could contribute to the ATIC signal (namely, PSR B0919+06, B0355+54, B1055+54, J1849-0317 and B1742-30, see fig.~\ref{fig:atic}), together with the objects, listed above, that might contribute to the PAMELA signal. We give the age, distance and the estimated $e^\pm$ energy output for these pulsars in tab.~\ref{tab:combination}. Guided by the analysis we carried out in the two previous sections, we want to contrast, here, a scenario where the PAMELA and ATIC data are accounted for by a single bright source, namely Geminga \cite{Yuksel:2008rf,Hooper:2008kg}, to a scenario where a combination of pulsars coherently contributes to the observed $e^\pm$ fluxes (tab.~\ref{tab:combination}).

\begin{figure}
\includegraphics[width=10cm,clip]{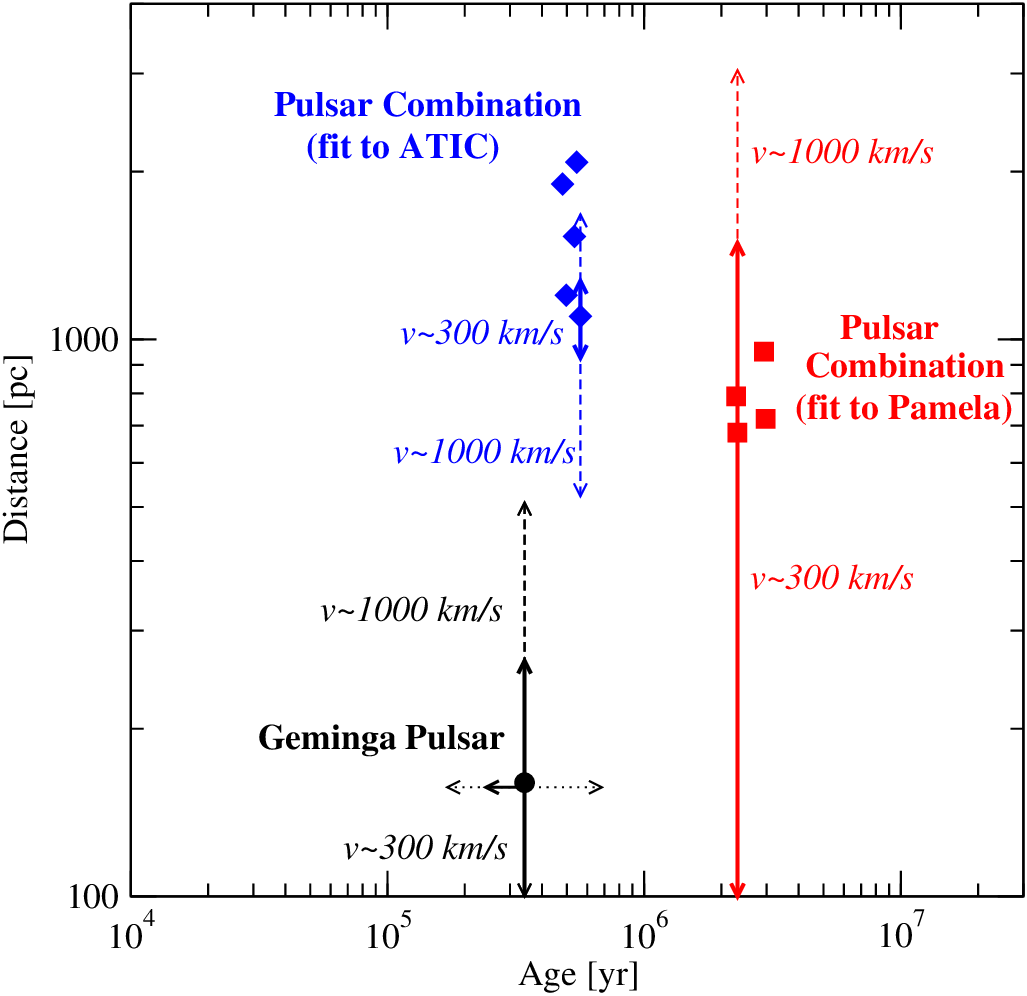}
\caption{\label{fig:bestfit_distage}The location, in the age versus distance plane, of the pulsars of tab.~\ref{tab:combination}. The vertical arrows indicate the potential, maximal effect of the pulsars' proper velocity, for a velocity of 300 km/s (solid lines) and of 1000 km/s (dashed lines). The horizontal arrows pointing outwards from the Geminga pulsar point indicate a range of a factor 2 uncertainty in the determination of the actual pulsar age from the characteristic age obtained from timing measurement (dotted lines) and the effect of the finite time it takes for the $e^\pm$ injected by the pulsar to diffuse in the ISM, assumed to be of order $10^5$ yr (solid line).}
\end{figure}

We adopt the MAX diffusion setup, and a spectral index $\alpha=1.85$ for all pulsars, including Geminga. In addition, we use the ST model with the reference positron-electron output efficiency factor $f_{e^\pm}=3\%$ for the combination of pulsars, and a lower efficiency factor for the single source Geminga, $f_{e^\pm}=0.45\%$. Also, following the estimate of Ref.~\cite{2004ApJ...601..340K}, we assume an exponential cutoff for the Geminga injection spectrum, at an energy of 0.67 TeV. It is of course a constraining assumption to take the same spectral index and the same positron-electron output efficiency for all pulsars in the combination we adopt. There is no reason to believe that every pulsar has the same value for these parameters. However, if we indeed find satisfactory agreement with data, this will reassure us and make our point stronger; had we not found good agreement, we could have massaged the individual pulsar parameters and gotten much better agreement.

We show in fig.~\ref{fig:bestfit_distage} the location  on the age versus distance plane of the pulsars reported in tab.~\ref{tab:combination}, which we use as combinations favored by observational data. Again, we point out that the ATIC data appear to be compatible with an origin from sources located at a distance of 1-2 kpc, and in a narrow range of ages, 0.4-0.6 Myr. PAMELA points towards closer (0.7-1 kpc) and older (2-4 Myr) pulsars instead. We also show the position of the Geminga pulsar. For the standard background we use the best GALPROP \cite{Strong:2001gh} self-consistent predictions for the primary and secondary electrons plus the secondary positrons, and we add a cutoff in the primary electron spectrum around 1 TeV, compatibly with the H.E.S.S. data of Ref.~\cite{Collaboration:2008aa}. 

In fig.~\ref{fig:bestfit_distage} we wish to draw again the attention of the reader to the uncertainties that affect the determination of the distance and age of pulsars when they actually inject high-energy cosmic-ray electrons and positrons in the ISM. In particular, the vertical arrows indicate the potential, maximal effect of the pulsars' proper velocity, for a velocity of 300 km/s (solid lines) and of 1000 km/s (dashed lines), values close to the average and maximal range of pulsar kicks \cite{1997MNRAS.291..569H}. The horizontal arrows pointing outwards from the Geminga pulsar point visually indicate, instead, a range of a factor 2 uncertainty in the determination of the actual pulsar age from the characteristic age obtained from timing measurement (dotted lines) and the effect of the finite time it takes for the $e^\pm$ injected by the pulsar to diffuse in the ISM, assumed to be of order $10^5$ yr (solid line).

\begin{figure*}
\mbox{\includegraphics[width=8.cm,clip]{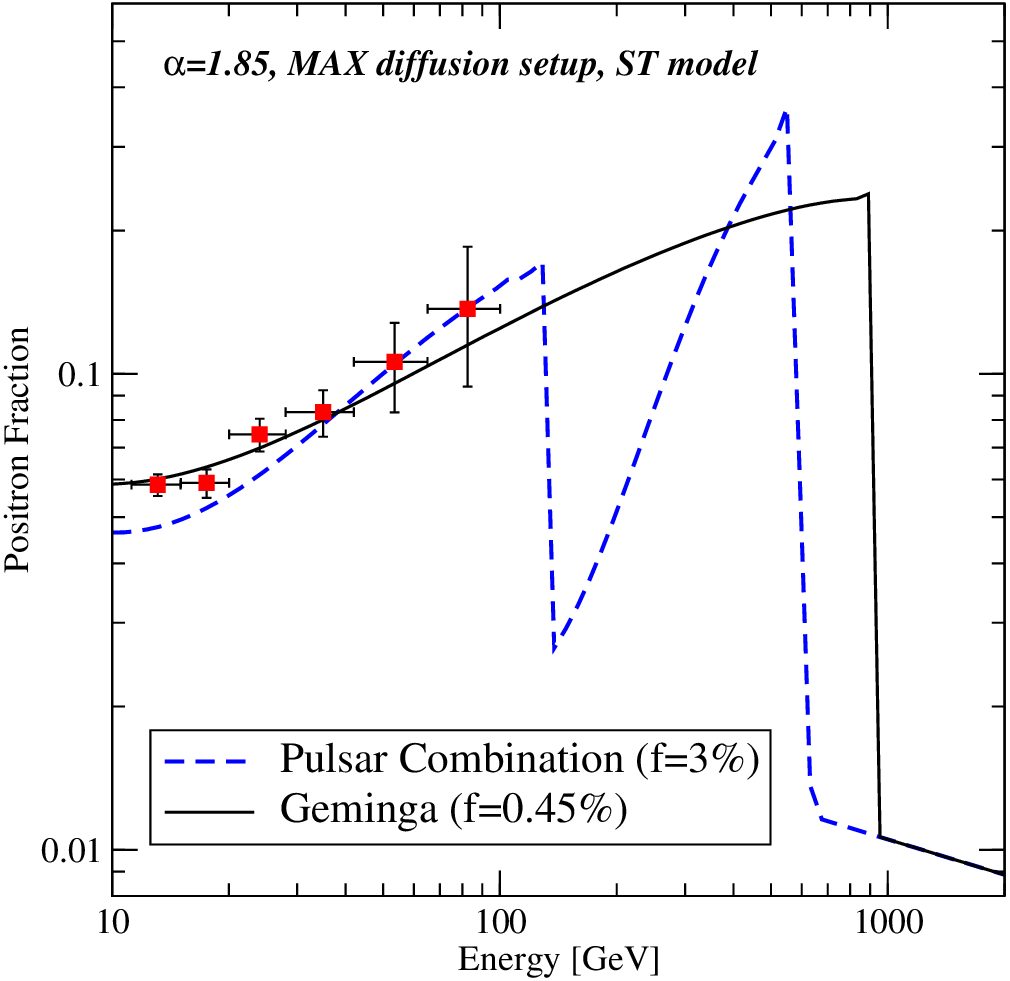}\qquad\includegraphics[width=8.cm,clip]{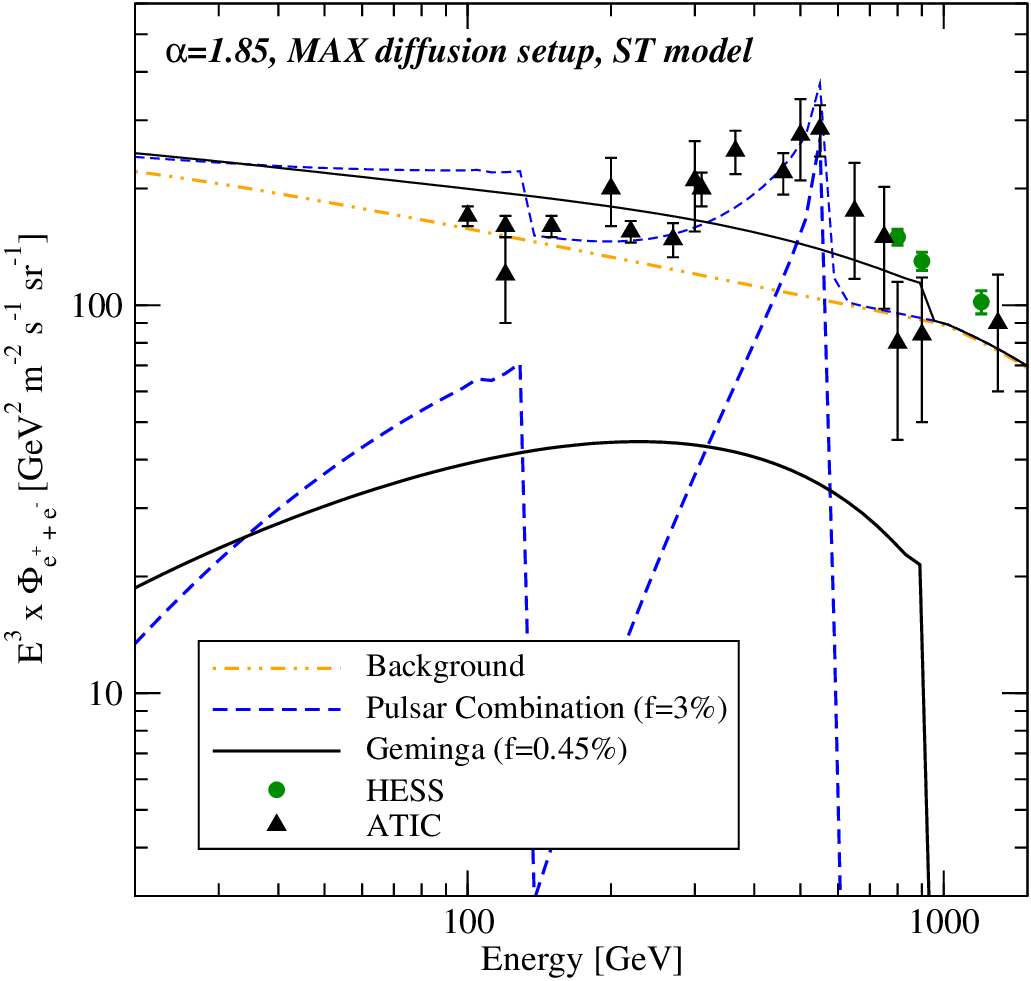}}
\caption{\label{fig:bestfit}The positron ratio (left) and the $e^\pm$ flux (right) for the Geminga pulsar (with $f_{e^\pm}=0.45$\%) and for the combination of multiple pulsars specified in tab.~\ref{tab:combination} (with $f_{e^\pm}=3$\%). We assume the MAX diffusion setup and an injection spectral index $\alpha=1.85$.}
\end{figure*}

The resulting positron fraction and overall $e^\pm$ differential energy spectrum for the combination of pulsars and for Geminga are shown in fig.~\ref{fig:bestfit}. As far as the PAMELA data are concerned, we obtain in both cases excellent fits. Again, had we chosen a non-trivial distribution of spectral indexes in the pulsar combination, the agreement could have been arbitrarily better. The positron fraction in the two cases (single versus combination) has vastly different features: The ``ATIC pulsars'' induce a sharp rise in the positron fraction after the abrupt cutoff of the ``PAMELA'' pulsars. The expected positron fraction therefore has a two-bumps structure. Of course, there is no reason not to assume that more pulsars would contribute at intermediate energies ($E_{e^\pm}\sim$100...200 GeV) and somewhat fill in the gap. In the case of Geminga, the prediction is a sharp drop around one TeV, and a consistent growing trend with energy below a TeV.

The right panel shows the overall $e^\pm$ spectrum in the two scenarios. Clearly, a single source does a rather poor job at fitting the detailed features in the ATIC data, while a combination of pulsars traces almost perfectly the feature reported by that experiment in the 600 GeV range. Notice that a combination of pulsars also shows a prominent drop-off feature at lower energies, which could be detected with higher statistics, e.g. by the Large Area Telescope (LAT) onboard Fermi (see sec.~\ref{sec:glast}).

In summary, the PAMELA data are perfectly compatible with both a single source and with a combination of  coherently contributing  pulsars. The ATIC data favor a scenario with more than one contributing source, and the resulting combination produces an interesting and very clean signature in the positron fraction.

In this section we considered only a sub-section of the numerous pulsars in the ATNF catalogue. It is important and crucial, however, to understand whether our predictions are consistent with adding up all other objects in the catalogue, and where and how the more than 1,000 other pulsars contribute to the production of energetic $e^\pm$. We address this items in the next section.

\section{Sanity Check: adding up all ATNF Sources}\label{sec:all}

\begin{figure*}
\mbox{\includegraphics[width=7.cm,clip]{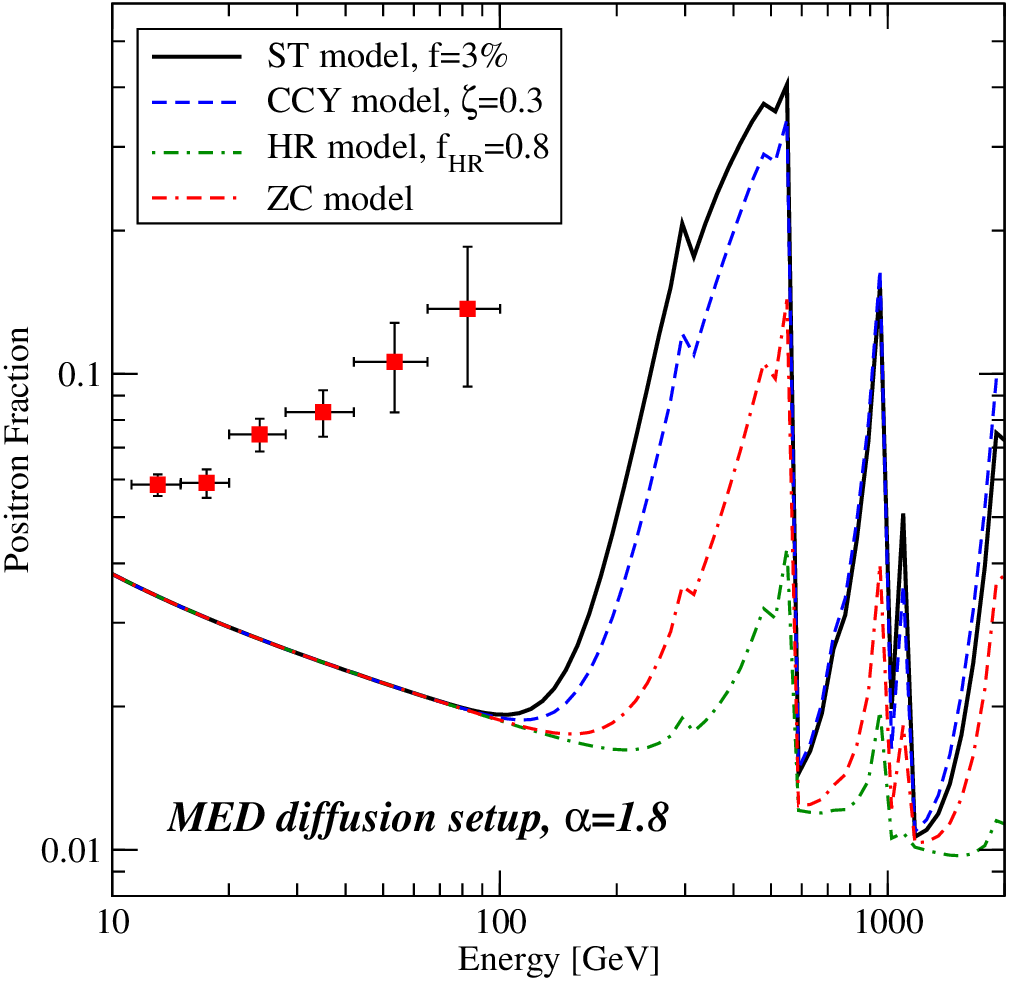}\qquad\includegraphics[width=7.2cm,clip]{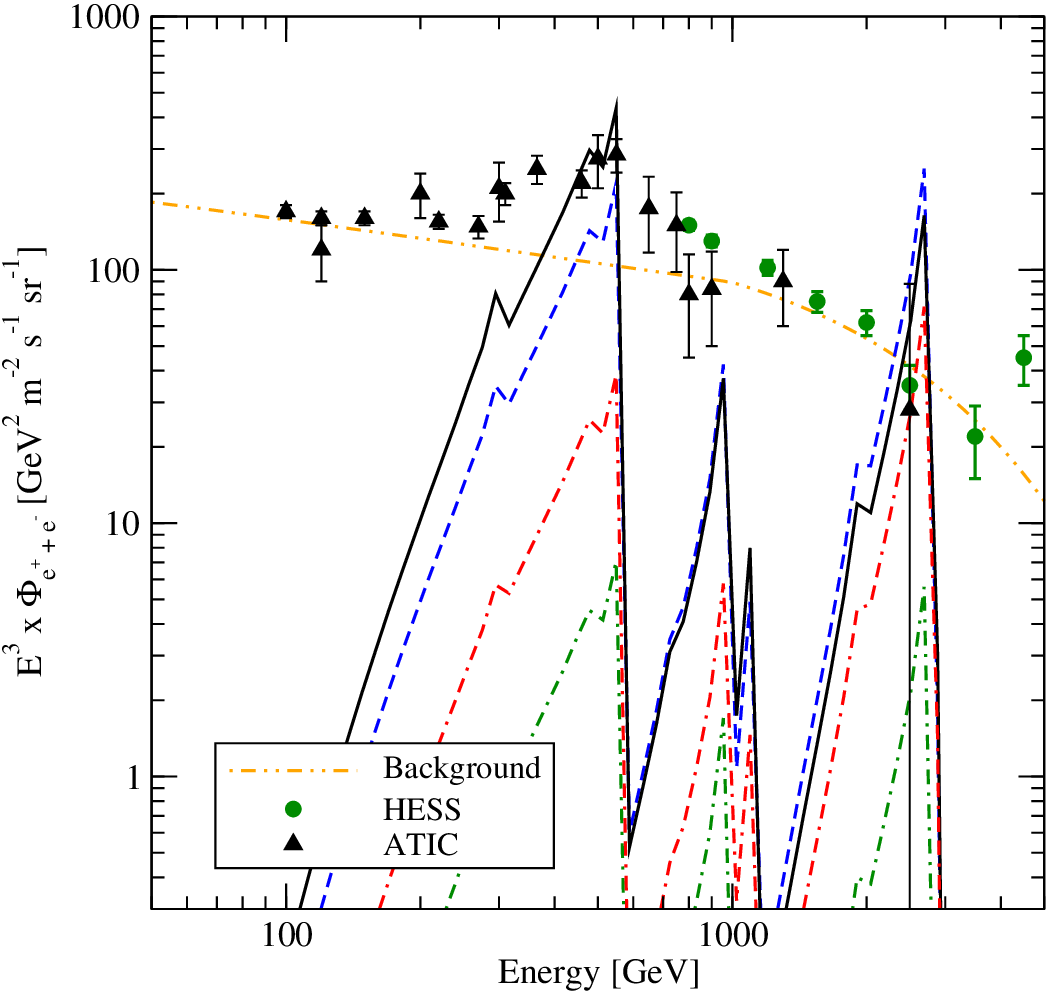}}\\[0.5cm]
\mbox{\includegraphics[width=7.cm,clip]{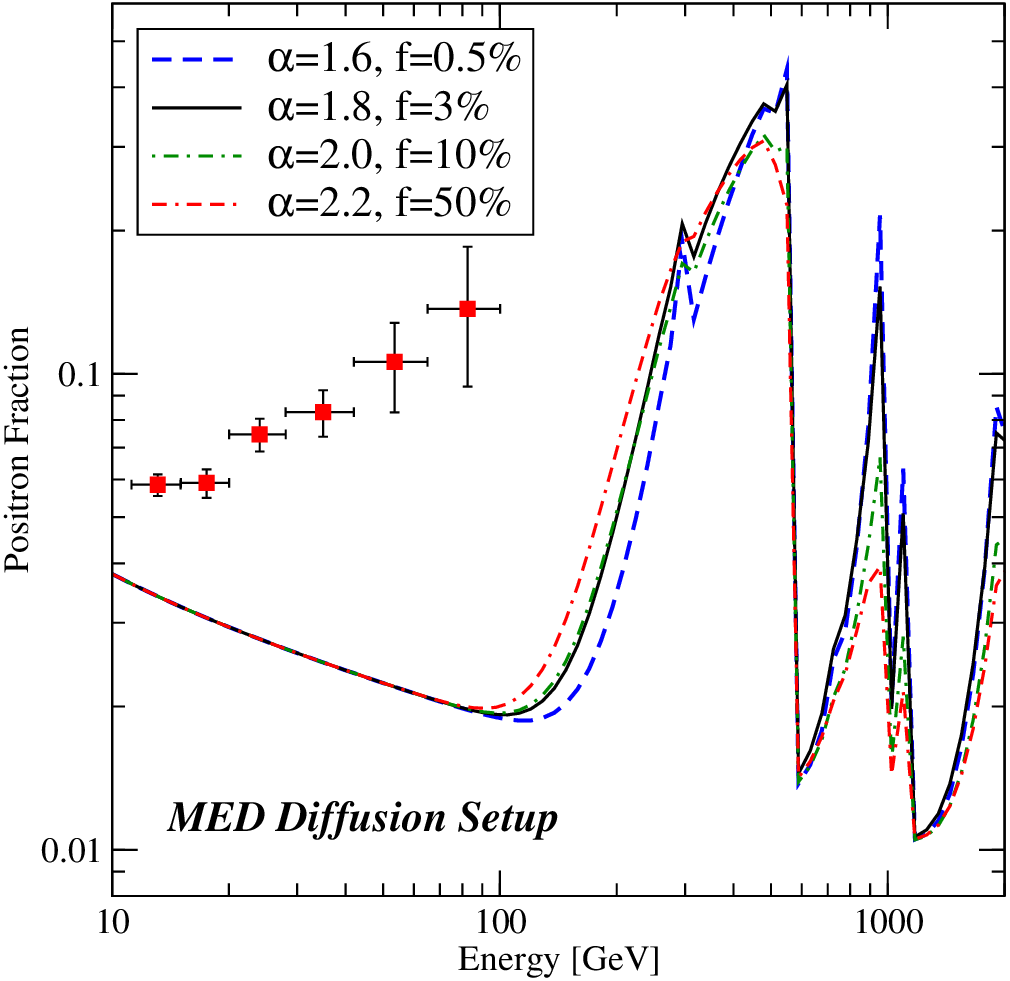}\qquad\includegraphics[width=7.2cm,clip]{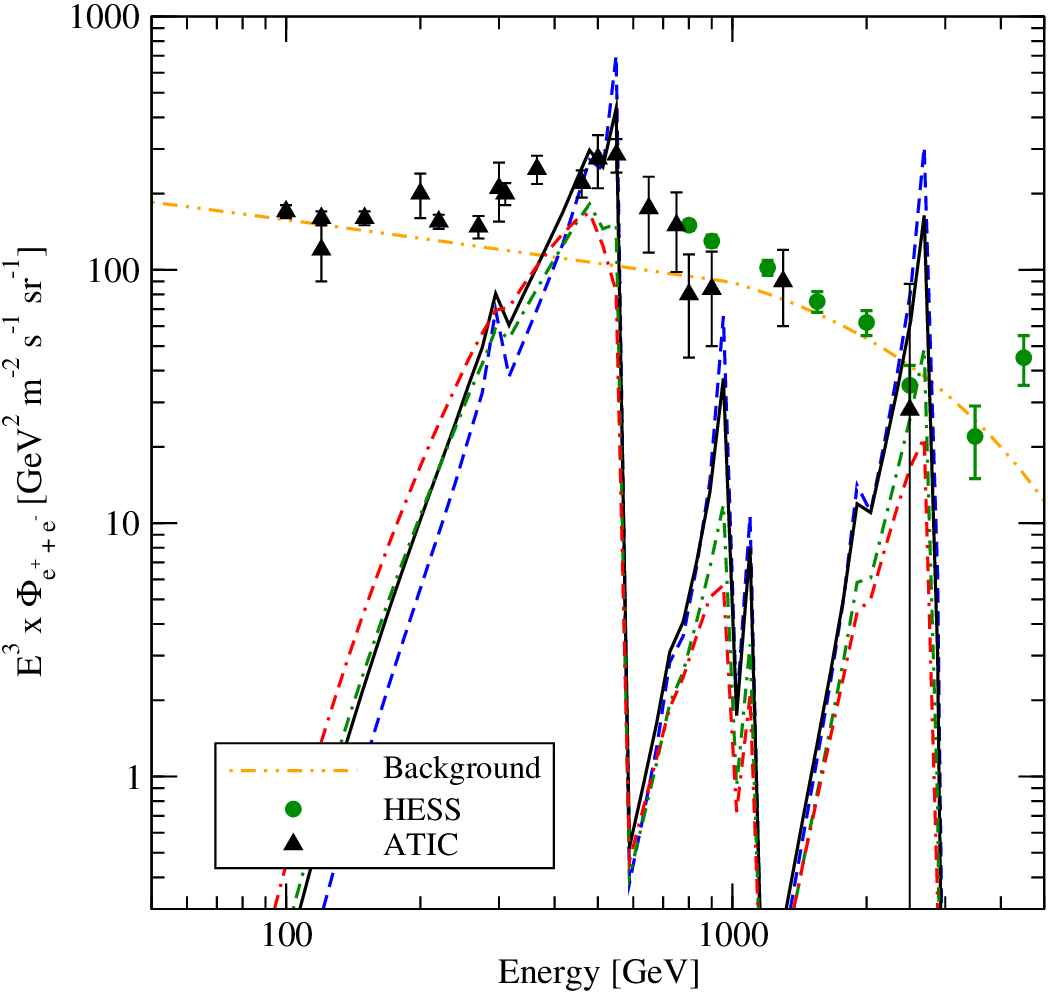}}\\[0.5cm]
\mbox{\includegraphics[width=7.cm,clip]{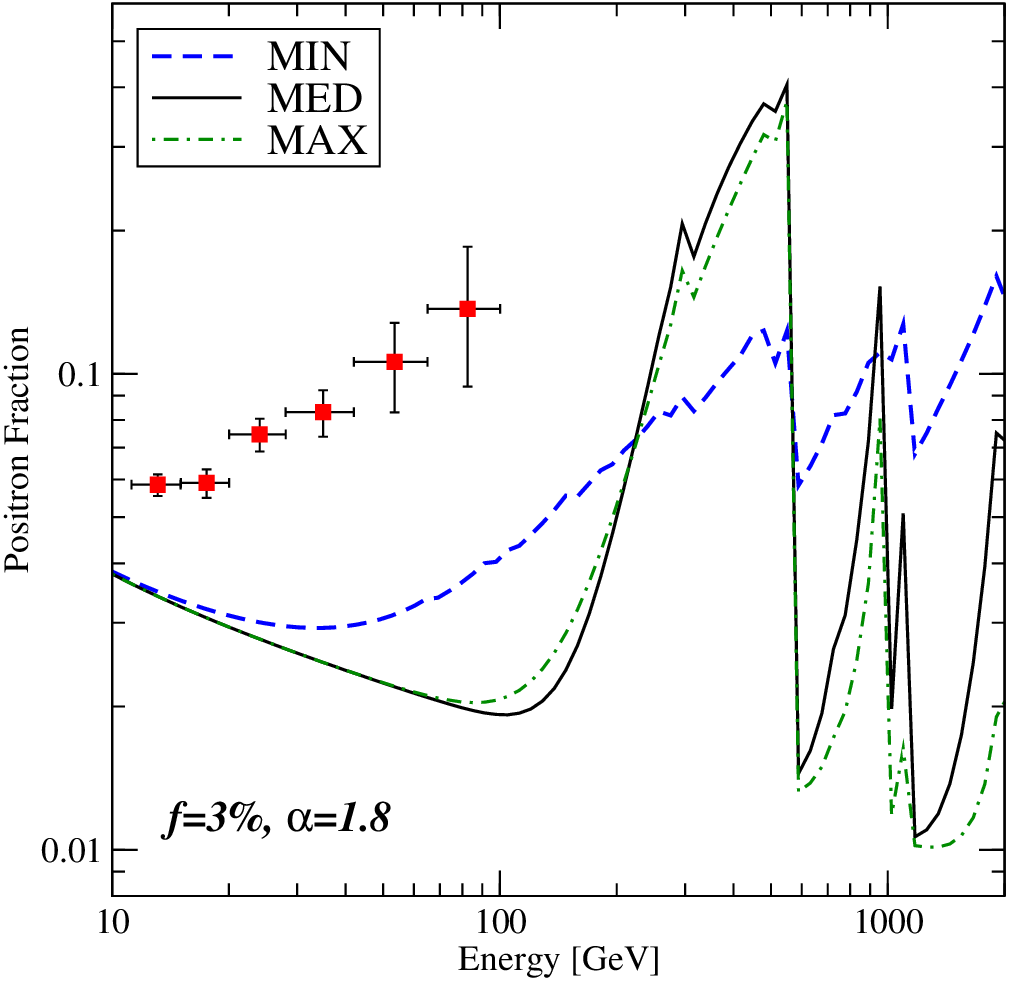}\qquad\includegraphics[width=7.2cm,clip]{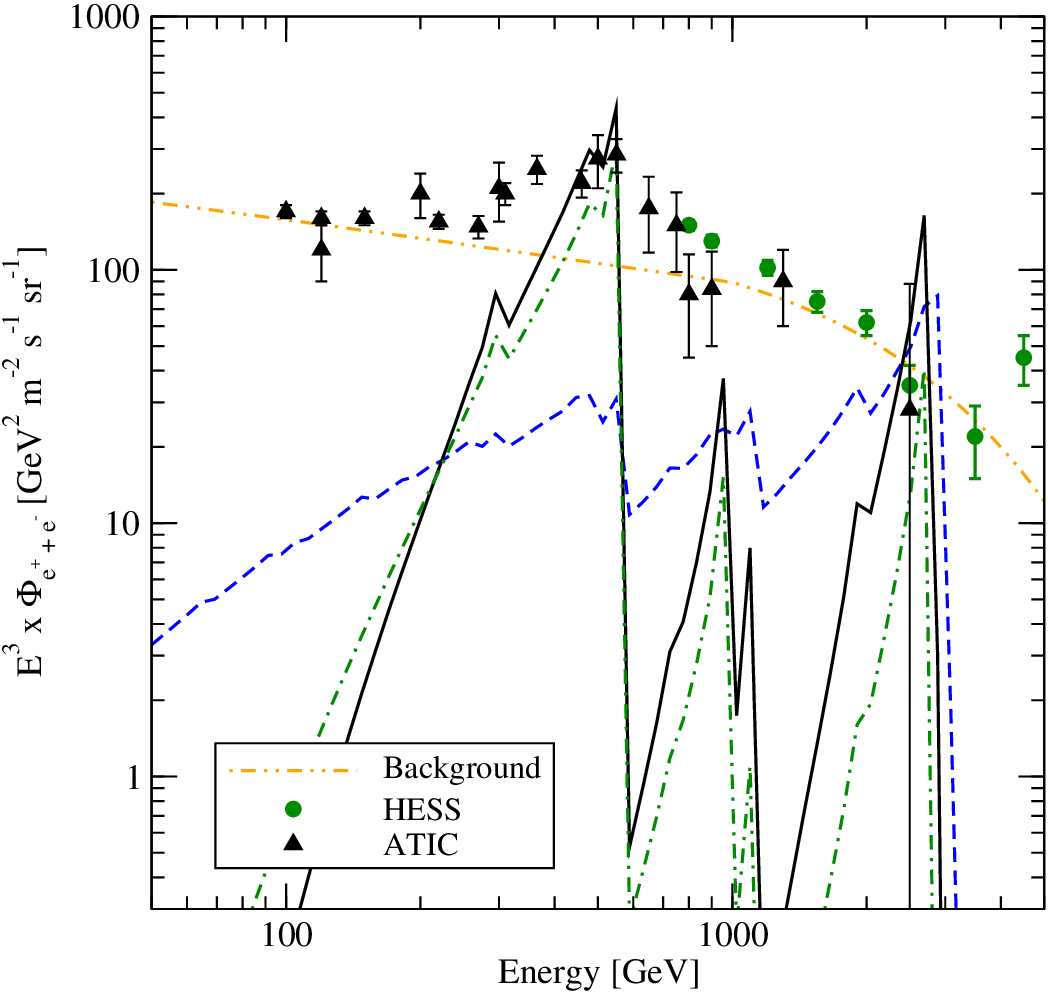}}
\caption{\label{fig:distant}The contribution from all gamma-ray pulsars ($g<1$) in the ATNF catalogue located at a distance greater or equal than 1 kpc. The left panels show the positron fraction, and the right panels the $e^\pm$ flux from the pulsars. In the top panels we study the variation in the contribution from relatively distant sources from the model with which we compute the $e^\pm$ output. In the middle panels we study how the contribution changes with the assumed spectral index, and in the bottom panels with the diffusion setup.}
\end{figure*}

We study in this section the overall contribution of all $g<1$ (``gamma-ray'') relatively distant pulsars (distances larger than 1 kpc) to the positron fraction and to the $e^\pm$ differential flux.
Overall, we include 248 objects, of the 266 total gamma-ray pulsars in the ATNF catalogue. The remaining nearby pulsars (distances less than 1 kpc) which are not included here have either been discussed in sec.~\ref{sec:nearby} or in sec.~\ref{sec:pamela}, \ref{sec:atic} and \ref{sec:onemulti}, or are too faint to give any significant contribution.

Scope of this section is to ascertain that within the models outlined in sec.~\ref{sec:models} for the pulsar $e^\pm$ energy output, existing pulsars do not over-produce energetic $e^\pm$. This exercise is therefore a ``sanity check'' to our approach. On the other hand, we will be able to appreciate what the contribution of existing, well-known pulsars is, for various diffusion models and average injection spectral index, and to decompose it in terms of distance and age.

While our approach is closer to observations than theoretical extrapolations based on assumed spatial distributions and pulsar birth rates \cite{2001A&A...368.1063Z,Hooper:2008kg}, an important point to keep in mind is the issue of {\em incompleteness}: as for all astrophysical catalogues, not all galactic pulsars have been detected yet  (and are therefore not included in the ATNF catalogue) within the age-distance domain we consider here. One aspect of the catalogue incompleteness is due to the simple fact that some fraction of the pulsars that could contribute to the local flux of $e^\pm$ are not bright enough to have been detected so far. A second aspect is that the pulsars radio emission is {\em beamed}, and depending on the angles between the pulsar axis of rotation, the direction of the beam and the observer pulsars can be radio-quiet. This source of incompleteness can be very relevant: for instance, Ref.~\cite{1998MNRAS.298..625T} estimates the fraction of pulsars which are actually beamed in the direction of an observer at Earth, as a function of the pulsar period in seconds, as $f(P)=0.09[\log(P/{\rm s}-1]^2+0.03$. In addition, the Reader should bear in mind that the same incompleteness from beaming might be true for other wavelengths, including X-ray and gamma-ray emission, although it is matter of debate how to estimate the incompleteness in those frequencies, and more specifically how this would impact those sources relevant for the present discussion. For sure, the Fermi Large Area Telescope (LAT) is giving, and will keep giving in the near future, a crucial contribution to the discovery of numerous new bright gamma-ray pulsars (potentially radio-quiet) that could very well contribute to the local flux of energetic  $e^\pm$ (see sec.~\ref{sec:glast}).

Fig.~\ref{fig:distant} shows the positron fraction (left panels) and the total $e^\pm$ flux for different models for the $e^\pm$ pulsar output (top panels), for the injection spectral index $\alpha$ (middle panels) and for the diffusion setup (bottom panels). As evident from the discussion in sec.~\ref{sec:models}, the ST and CCY models give comparable $e^\pm$ outputs, with the latter boosting the output of more nearby pulsars (those contributing to the highest $e^\pm$ energies). The ZC and the HR model predict an almost negligible contribution to both the positron fraction and the overall $e^\pm$ flux from distant pulsars. In general, in all scenarios, the positron fraction starts to increase after 100 GeV (unless one assumes a MIN diffusion setup), so it is unlikely that the PAMELA positrons dominantly originate from further away than 1 kpc.

The spiky shape of the curves of fig.~\ref{fig:distant} should be interpreted with care: the reader should keep in mind that we are assuming a hard spectral index, and no cutoff in the power-law injection spectrum at high energies. In addition, we neglect the temporal profile of the $e^\pm$ injection, which again would smooth out the spikes we find. Lastly, the finite energy resolution of any $e^\pm$ detector would also make the spikes appear much smoother in the actual data.

The middle panels show that no significant qualitative variation is induced by changing the $e^\pm$ injection spectral index from $\alpha=1.6$ to 1.8, 2.0 and 2.2. The harder the spectral index, the steeper the spikes predicted in both the positron fraction and in the overall differential spectrum.

A more significant role is played by the diffusion setup. The MIN model drives energetic $e^\pm$  to lower energies much more effectively than the MAX and MED models, which enhances the contribution of distant pulsars to positrons in the PAMELA energy range. The MED and MAX setups, on the other hand, give comparable predictions for the contribution of distant pulsars.

\begin{figure*}
\mbox{\includegraphics[width=8.cm,clip]{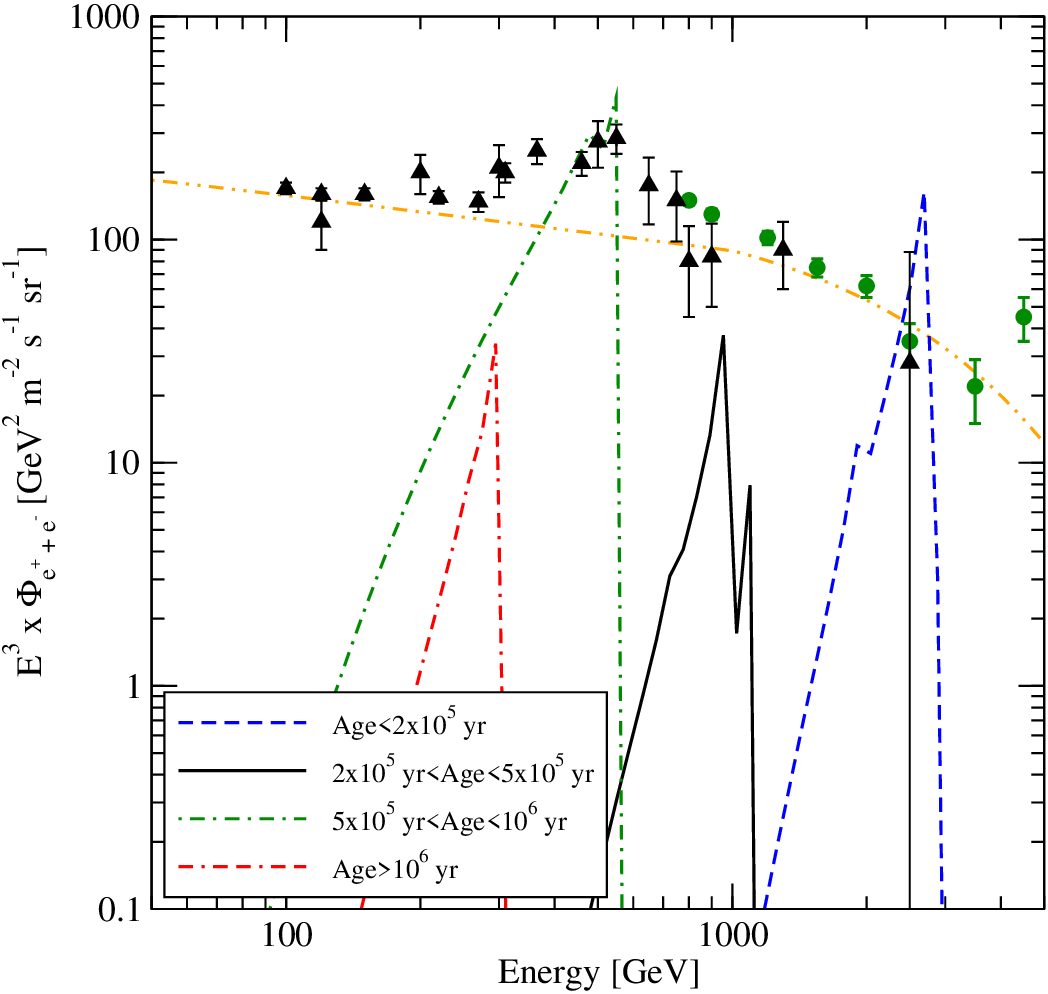}\qquad\includegraphics[width=8.cm,clip]{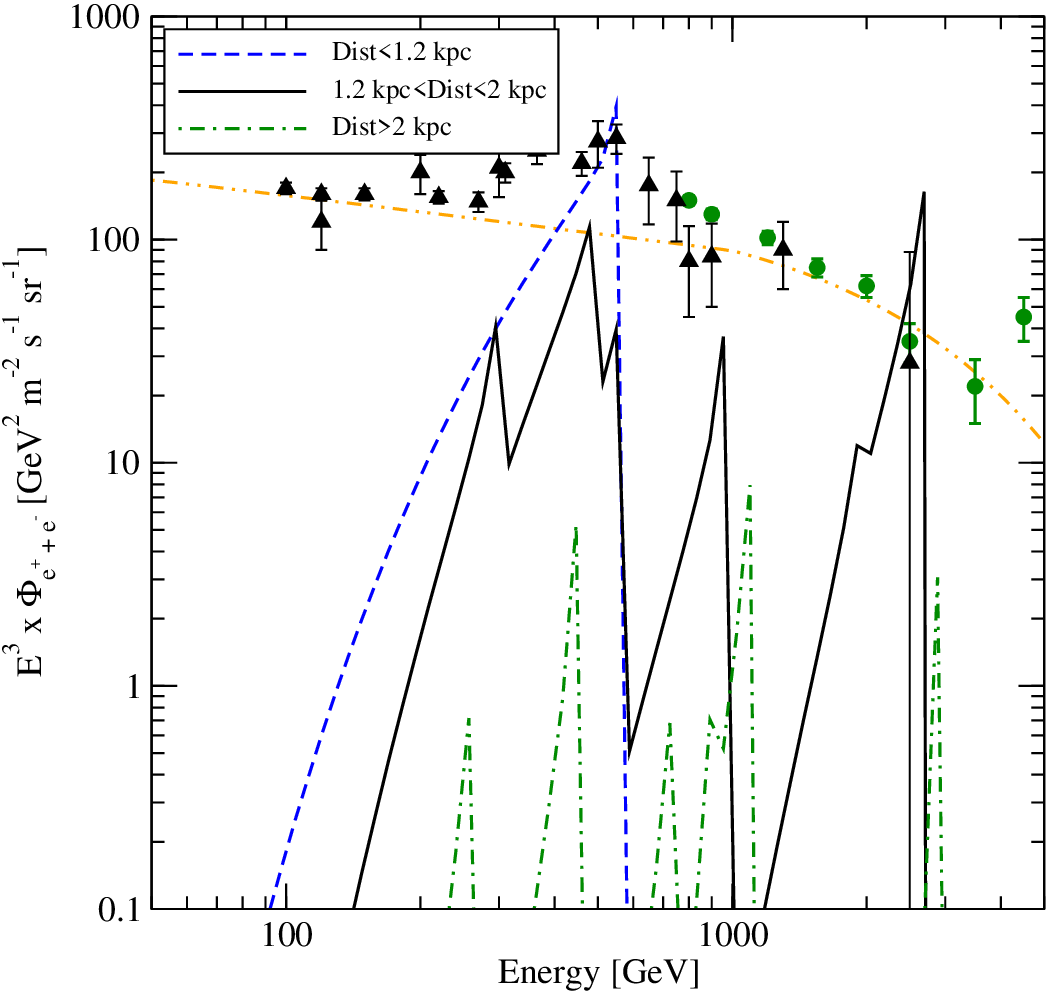}}
\caption{\label{fig:distageall}Break-up of the contribution from distant ($>1$ kpc) pulsars in age (left) and distance (right), for $\alpha=1.8$ and the MED diffusion setup.}
\end{figure*}

In fig.~\ref{fig:distageall} we break up the contribution from distant pulsars to the $e^\pm$ spectrum in intervals of age (left) and distance (right), assuming a MED diffusion setup, the ST model and $\alpha=1.8$. As expected, the youngest pulsars contribute to the spike at large energies; pulsars with ages between 0.2 and 0.5 Myr contribute dominantly around 1 TeV, while in the range of the ATIC bump the bulk of the contribution from distant pulsars stems from those objects with an age between 0.5 and 1 Myr, as already observed above. Finally, older pulsars give a subdominant contribution at even lower energies.

The break-up with distance highlights that the ATIC bump is plausibly mostly dominated by the nearest objects, with some contribution from pulsars as distant as 2 kpc. Sources further than 2 kpc give a negligible contribution, spread out through the entire energy spectrum, given the possible age range.

The bottom line of this section is that the models we used  for the computation of the pulsars' $e^\pm$ output are consistent with all relevant objects in the ATNF catalogue (pulsars with $g>1$ give an even smaller contribution), and distant sources, while unimportant or subdominant for the PAMELA energy range, might indeed be relevant to account for the spectral feature observed by ATIC.

\section{The Role of Fermi-LAT: Spectral Measurement and New Pulsars}\label{sec:glast}

\begin{figure*}
\includegraphics[width=15.cm,clip]{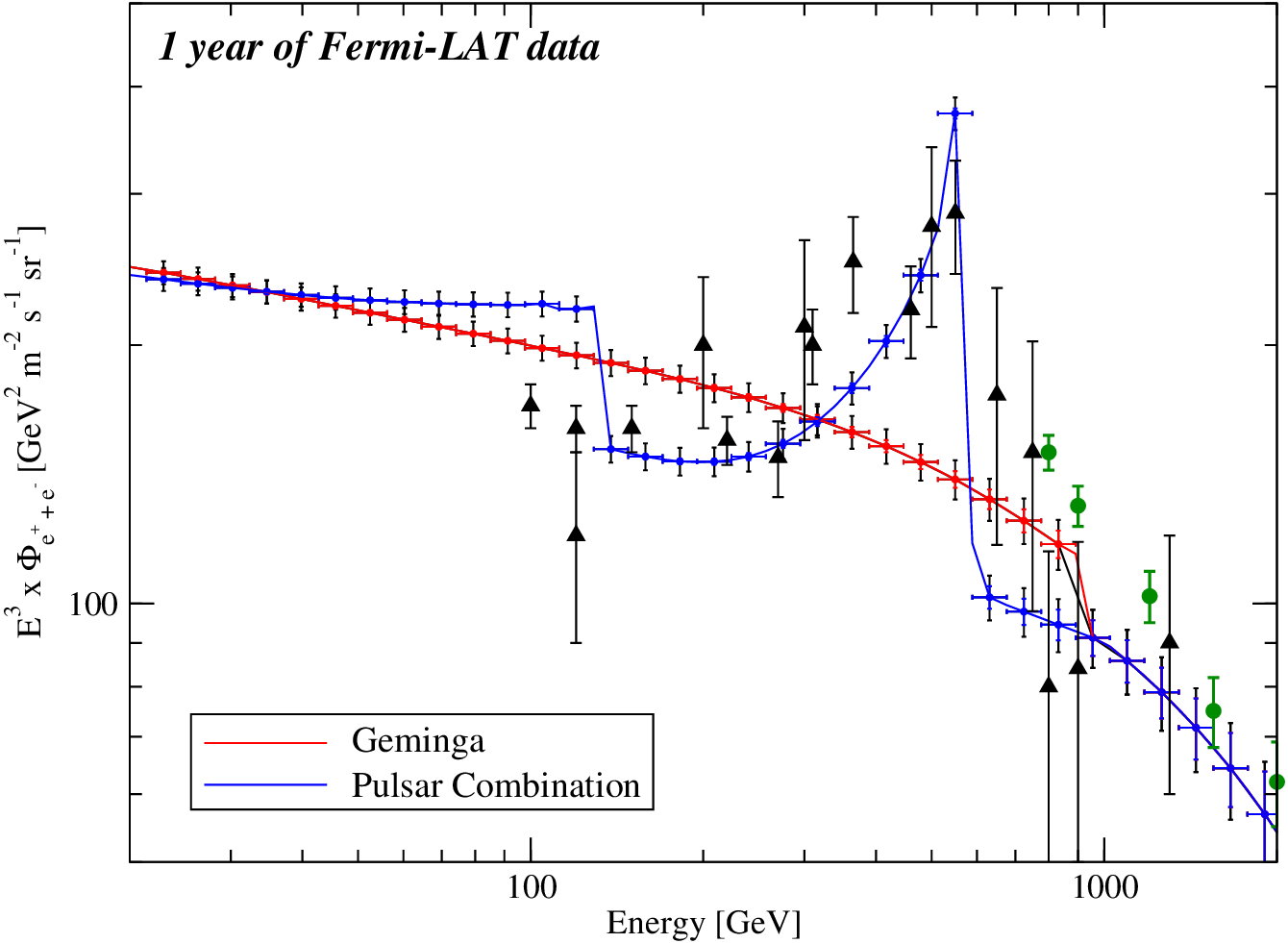}
\caption{\label{fig:glast}The anticipated data on the $e^\pm$ spectrum from one year of full science operation with the Fermi-LAT telescope, for the two cases, shown in fig.~\ref{fig:bestfit}, of one bright nearby source (Geminga) and of multiple, more distant pulsars accounting for both the PAMELA and the ATIC data.}
\end{figure*}

It has been recently pointed out that the Large Area Telescope (LAT) on board the Fermi Space Telescope can efficiently detect energetic cosmic ray electrons and positrons \cite{2007arXiv0706.0882M,2007AIPC..921..500M,2007arXiv0711.3033M}. Being a gamma-ray telescope based on pair-production, the LAT is, intrinsically, a high-energy electron and positron detector. The main problem one faces in measuring the $e^\pm$ spectrum with Fermi-LAT is to efficiently separate electrons and positrons from all other cosmic-ray species, mainly protons. The LAT team has successfully argued and demonstrated that this is actually possible, with an appropriate choice of analysis cuts. In the energy range between 20 GeV and 1 TeV, the effective geometric factor (for electrons), after applying the mentioned cuts, ranges from 0.2 to 2 ${\rm m}^2$ sr (enormously larger than both ATIC's and PAMELA's, especially when one additionally factors in the data-taking time as well), and the energy resolution is 5 to 20\%, depending on the energy. Applying the selection cuts
to the simulated cosmic ray flux, the residual hadron contamination was estimated to be $\approx 3$\% of the $e^\pm$ flux  \cite{2007arXiv0706.0882M,2007AIPC..921..500M,2007arXiv0711.3033M}. We use this number to evaluate the systematic error on the flux measurement in addition to statistical uncertainties.

Fig.~\ref{fig:glast} illustrates the capabilities of Fermi-LAT to measure the electron-positron spectrum  in one year of full science mode data taking, and assuming a geometric factor of 1 ${\rm m}^2$ sr. The energy bins are logarithmically evenly spaced, but they fall in the ballpark of the anticipated detector energy resolution ($\Delta E_{e^\pm}/E_{e^\pm}\sim5...20$ \%). To estimate the error bars we linearly sum the systematic uncertainty from hadronic contamination and the statistical uncertainty (the former in black, the latter in blue or red, and assumed to be normally distributed, and almost invisible at low energies). Systematic errors dominate at low energies, while at larger energies the $e^\pm$ counts get low enough that statistical errors become dominant. Both statistical and systematic errors will improve with more data taking time and with a better understanding of the detector. The take-away message is, though, that Fermi will provide an exquisite measurement of the $e^\pm$ spectrum. In the figure we use the two models of sec.~\ref{sec:onemulti} for illustration. It is clear that in both cases Fermi-LAT will have the sensitivity to (indirectly) observe a drop-off in the positron fraction (blue curve), if this occurs where predicted in our simple scenario that fits the PAMELA observations, and will definitely solidly confirm or refute the spectral feature observed by ATIC.

\begin{figure}
\includegraphics[width=14.cm,clip]{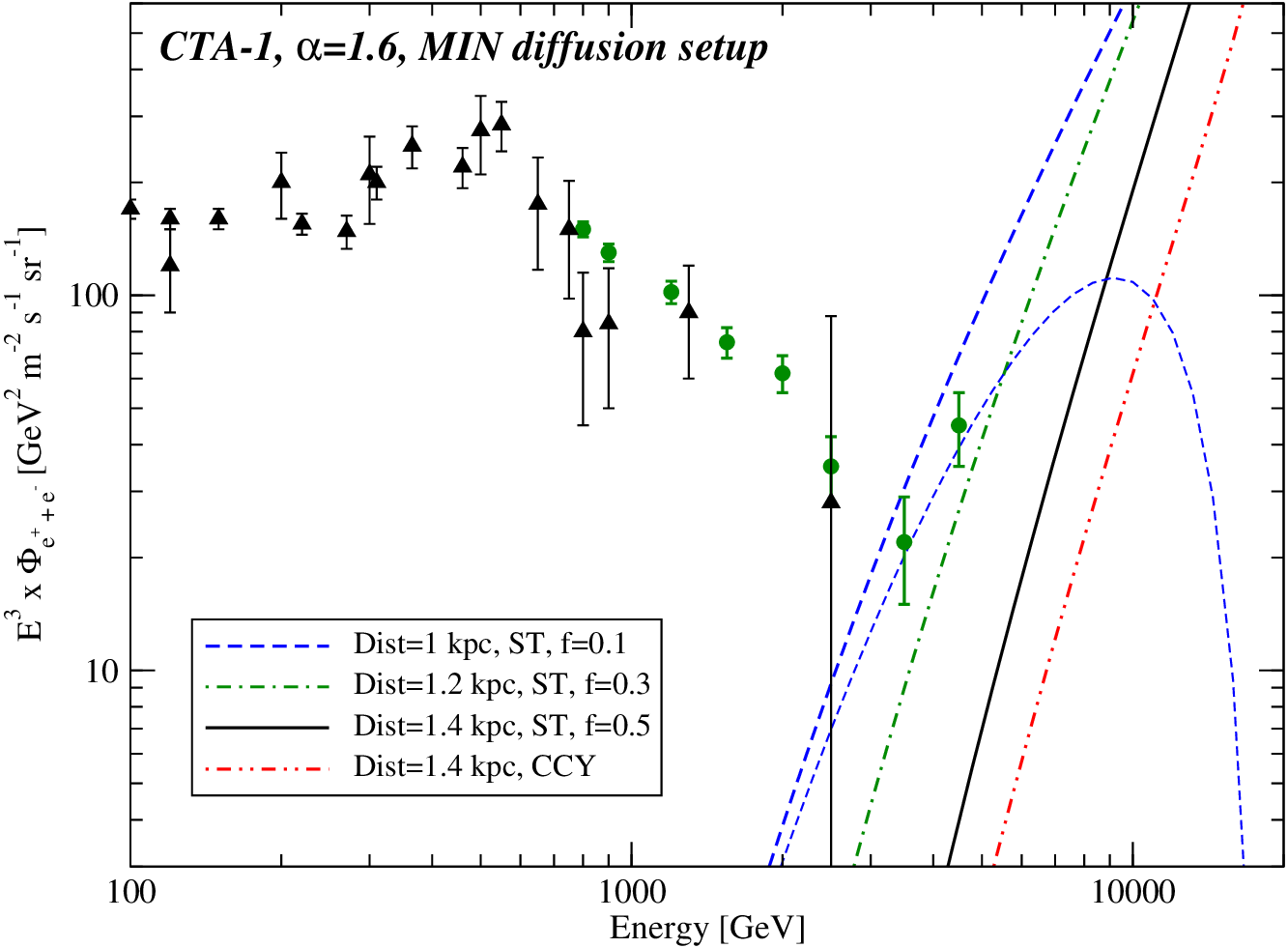}
\caption{\label{fig:CTA1}Predictions for the contribution to the total $e^\pm$ flux of the CTA1 pulsar, recently discovered by Fermi-LAT \cite{2008Sci...322.1218A}. We assume the MIN diffusion setup and a spectral index $\alpha=1.6$, as inferred from the EGRET data on the same gamma-ray source \cite{1998MNRAS.295..819B}. The thinner blue dashed line corresponds to the prediction for a cutoff at 10 TeV in the injection spectrum.}
\end{figure}

Fermi-LAT will play a second decisive role in understanding the origin of energetic $e^\pm$: the discovery of new gamma-ray pulsars. In addition to monitoring portions of the gamma-ray sky known to host radio-pulsars, the LAT Collaboration is actively and successfully carrying out a program of blind search for gamma-ray pulsars, along the lines outlined in Ref.~\cite{Ziegler:2007zzb}. The first Fermi-LAT science paper was in fact about a newly discovered very bright gamma-ray pulsar in the CTA 1 region \cite{2008Sci...322.1218A}.  Very accurate measurements of the period and of its derivative made it possible to derive a rather accurate estimate of the age of the CTA 1 pulsar, which was found to be $T\simeq1.4\times 10^4$ yr \cite{Ziegler:2007zzb}. This age determination is compatible with the radio and X-ray pulsation of the objects, which also inform us on it having a distance of $1.4\pm0.3$ kpc. In addition, the LAT data indicate a rather powerful spin-down luminosity of $4.5\times 10^{35}$ erg/s, and a surface magnetic field $B_{12}\simeq11$.

Pulsars like CTA 1 are likely to contribute significantly to the $e^\pm$ flux. To illustrate this point, we plot in fig.~\ref{fig:CTA1} the predictions for the overall $e^\pm$ differential spectrum expected from the CTA 1 pulsar, for three values of the distance (1, 1.2 and 1.4 kpc) and the ST model (with values of $f_{e^\pm}=0.1$, 0.3 and 0.5, respectively), as well as for the CCY model, assuming a distance of 1.4 kpc. Consistently with the EGRET spectrum of the same source \cite{1998MNRAS.295..819B} we assume $\alpha=1.6$. We employ in this plot a MIN diffusion setup, and we show with the thin dashed blue lined the effect of an $e^\pm$ injection energy cutoff at $E_{e^\pm}\simeq10$ TeV.  Given that CTA 1 is a relatively young object, the contribution peaks at very high energy (at or more than 10 TeV). If the source is located between 1 and 1.2 kpc, it plausibly contributes to the highest energy bins in the H.E.S.S. data.  Fermi LAT has therefore the ability to discover numerous sources which are bright enough to give a significant contribution to the $e^\pm$ spectrum. 

Besides a very accurate measurement of the high-energy cosmic-ray lepton spectrum and the discovery of new local, bright pulsars, Fermi-LAT data will also provide directional information on high-emergy electrons and positrons. Although charged cosmic ray leptons are isotropize quickly in the ISM magnetic fields, if they are produced close enough to Earth, and with large enough energies, a potentially detectable dipolar anisotropy might be detected from the direction of nearby pulsars, if indeed the latter source the bulk of the local high-energy cosmic ray leptons (on this point see also Ref.~\cite{2004ApJ...601..340K} and the recent estimates of Ref.~\cite{Hooper:2008kg} and \cite{Buesching:2008hr}).

\begin{figure}
\includegraphics[width=14.cm,clip]{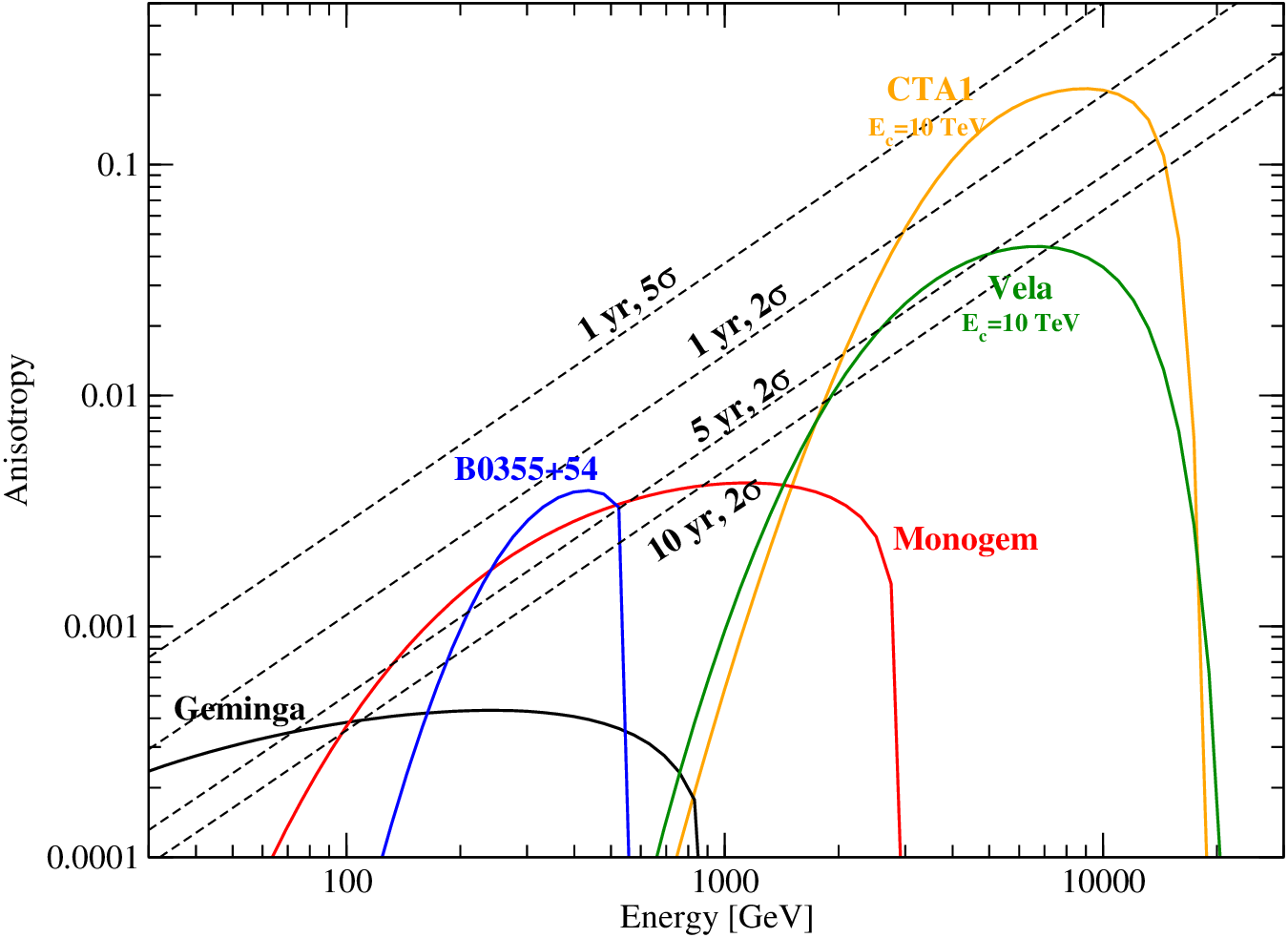}
\caption{\label{fig:anisotropy}Predictions for the anisotropy in the arrival direction of cosmic ray leptons from nearby pulsars, defined as $(I_f-I_b)/(I_f+I_b)$, where $I_f$ is the total number of events from the emisphere in the direction of the pulsar, and $I_b$ that from the opposite emisphere. The parameters are the same as in fig.~\ref{fig:spectrum_nearby} and \ref{fig:CTA1}, blue thin dashed curve for CTA1. For Vela and CTA1 we consider a cutoff at 10 TeV in the injection spectrum.}
\end{figure}

We show in fig,~\ref{fig:anisotropy} the predicted anisotropy (defined as usual as$(I_f-I_b)/(I_f+I_b)$, where $I_f$ is the total number of events from the emisphere centered on the direction of the pulsar, and $I_b$ that of events from the opposite emisphere), as a function of energy, produced by the pulsars we considered in sec.~\ref{sec:nearby} and in the present section. Specifically, we adopt here the same parameters as those employed in fig.~\ref{fig:spectrum_nearby} for Monogem, Geminga, B0355+54 and for Vela (for the latter we assume a cutoff at 10 TeV in the injection spectrum), and, for CTA1, the setup we employed for the thin dashed blue line (distance of 1 kpc, ST model with $f_{e^\pm}=0.1$, and a cutoff at 10 TeV in the injection spectrum). In the Fermi-LAT range, the largest anisotropies lie between one part in a thousand and one in a hundred, for the pulsars and the particular setup we employ here. Notice that the distance of B0355+54 suppresses the resulting anisotropy as opposed to a nearby object like Monogem, when comparing fig.~\ref{fig:anisotropy} with the fluxes shown in fig.~\ref{fig:spectrum_nearby}. 

No dedicated studies of the performance of Fermi-LAT in discerning an anisotropy in the arrival direction of high-energy cosmic-ray leptons exist yet, so it is hard to estimate whether anisotropies at this level might in fact be detected. Naively, one can conservatively estimate the number of events from the diffuse cosmic ray lepton background alone as a function of energy, and compare, in a given emisphere, a 2 or 5$\sigma$ poissonian fluctuation in that number to the anisotropy to be measured. We show curves, obtained in this way, corresponding to 1 yr of data, and to 2 and 5$\sigma$, as well as two 2$\sigma$ curves for 5 and 10 yr.

The best chances Fermi-LAT appears to have to detect one of the bright nearby pulsars is (i) for a mature and nearby object like Geminga, in the low energy part of the spectrum; (ii) for a younger, but farther bright pulsar like Monogem, in the several hundred GeV range, or (iii) for very young pulsars, like Vela or CTA1, in the multi-TeV range, if the systematics can be controlled enough to reconstruct with accuracy electron events at those energies. This latter case can be in principle nicely complemented with data from ACT's as well. In any case, it appears that an anisotropy study of the Fermi-LAT is potentially feasible and even promising.

We point out, however, that while the detection of an excess from the direction of a nearby pulsar would be a very suggestive experimental result, the absence of an anisotropy feature at the level predicted here would not be a conclusive evidence {\em against} the pulsar scenario (for instance due to local, solar system effects, or to the large scale structure of ISM magnetic fields, or because of the proper motion of the pulsars themselves, and thus uncertainties in the actual site of $e^\pm$ injection).

To conclude, we wish to stress here that the Fermi telescope will play an essential role in forming a {\em self-consistent picture} of local high-energy cosmic ray electrons and positrons, specifically in connection with, but not limited to, the pulsar scenario.

\section{Summary and Conclusions}\label{sec:concl}

We argued in this study that there is indeed no ``need to add entities beyond necessity'' to explain the PAMELA and the ATIC data on cosmic-ray electron and positron fluxes: existing and well-known 
pulsars easily and naturally (from an energy budget standpoint) account for both experimental results.

We summarize below the main results of the present theoretical study.
\begin{itemize}
\item Local pulsars are well-known sites of electron-positron pair production. Although a firm theoretical understanding of the pulsars' $e^\pm$ energy output is lacking, we surveyed here a few possibilities, which are all approximately in line with one another and with estimates of the output energy in electrons from SNR's. Remarkably, by picking a few nearby known pulsars and employing the mentioned energy output models, we easily reproduce the spectral features and intensities reported by both PAMELA and ATIC, with nominal values for the $e^\pm$ output efficiency;
\item The quality of the recent data on the positron fraction and on the differential $e^\pm$ flux is good enough to allow the exploration of the regions of the pulsar parameter space favored by the two experimental observations; uncertainties in the determination of pulsars' ages and distances -- particularly, as far as the latter are concerned, stemming from the large observed pulsar velocities -- can however significantly compromise our ability to pinpoint single major contributors to the local cosmic-ray lepton flux
\item Although statistically and energetically plausible for existing objects, we find that the PAMELA and ATIC data are not well fitted by one single bright nearby source;
\item For a given $e^\pm$ injection spectrum and diffusion setup, the PAMELA data point towards a corridor of correlated values of pulsar age and distance. Several known pulsars fall within this corridor, and could coherently add up to contribute to the positron excess over the standard secondary population observed by PAMELA;
\item The ATIC feature at around 600 GeV can also be explained by one or more existing pulsars, although this generically requires a rather high electron-positron energy output, and, unlike PAMELA, it does not seem to be a natural or necessary {\em post}-diction of the pulsar contribution to energetic $e^\pm$: while it is plausible to expect pulsars to produce a positron fraction like that measured by PAMELA, it is not as obvious, in the same context, to expect a spectral feature like that reported by ATIC. Nevertheless, relatively young (0.4-0.6 Myr) pulsars 1-2 kpc away can easily account for a spectral feature such as the one observed by ATIC.
\item The Fermi Space Telescope will play an extraordinarily important role in the understanding of the local flux of energetic electrons and positrons, by (1) providing us, in the very near future, with an exquisite measurement of the $e^\pm$ flux, that could unveil several interesting features, and/or rule out recent experimental claims, and (2) by discovering gamma-ray pulsars that can add-up to give a self-consistent picture of the origin of the local energetic electrons and positrons, and (3) searching for anisotropies in the arrival direction of high-energy electrons and positrons.
\end{itemize}
A convincing picture of the nature of energetic cosmic ray electrons and positrons evidently calls for a two-pronged effort: on the one hand, theorists working on cosmic rays and on compact objects should refine the theoretical understanding of $e^\pm$ production in pulsars and possibly in other astrophysical galactic objects; on the other hand, further and deeper observations of candidate electron-positron sources at various wavelengths (including radio, X-ray, gamma-ray and very high energy gamma-ray frequencies) will be needed to complement and to drive the theoretical effort towards the construction of a tho\-rough\-ly satisfactory picture.


  
\begin{acknowledgments}
This work originated from an invited talk on ``Astrophysical interpretations of the positron excess'' at the ``New Paradigms for Dark Matter'' workshop at  the University of California, Davis, December 5-6, 2008. I am indebted to many discussions with the workshop participants. I also acknowledge useful discussions and inputs from Michael Dine, Marc Kamionkowski and from several members of the Fermi Collaboration, in particular Elliott Bloom, Tesla Jeltema, Alex Moiseev and Robert Johnson. We also gratefully acknowledge the very detailed comments and suggestions we received from an anonymous Referee.
This work is partly supported by US Department of Energy Contract DE-FG02-04ER41268 and by an Outstanding Junior Investigator Award, by NASA Grant Number NNX08AV72G and by NSF Grant PHY-0757911.
\end{acknowledgments}


\bibliography{pulsars_resub}


\end{document}